\documentclass[a4papers]{cas-sc}

\usepackage{graphicx}
\usepackage{lineno,hyperref}
\usepackage{color}
\usepackage{multirow}
\usepackage[authoryear,longnamesfirst]{natbib}
\modulolinenumbers[5]

\def\tsc#1{\csdef{#1}{\textsc{\lowercase{#1}}\xspace}}
\tsc{WGM}
\tsc{QE}

\begin{document}
\let\WriteBookmarks\relax
\def\floatpagepagefraction{1}
\def\textpagefraction{.001}

\shorttitle{
Structured jet model for GRB~080710
}   

\shortauthors{Obayashi et al. 2023}


\title[mode = title]{
GRB~080710: A narrow, structured jet showing a late, achromatic peak in the optical and infrared afterglow?
}

\author[1]{Kaori~Obayashi}
\ead{o-kaori@phys.aoyama.ac.jp}

\author[1]{Ayumu~Toriyama}

\author[1]{Mayu~Murakoshi}

\author[1]{Yuri~Sato}

\author[1,2]{Shuta~J.~Tanaka}

\author[1]{Takanori~Sakamoto}

\author[1,3]{Ryo~Yamazaki}

\address[1]{Department of Physical Sciences, Aoyama Gakuin University,
5-10-1 Fuchinobe, Sagamihara, Kanagawa 252-5258, Japan}
\address[2]{Graduate School of Engineering, Osaka University, 2-1 Yamadaoka, Suita, Osaka 565-0871, Japan}
\address[3]{Institute of Laser Engineering, Osaka University,
2-6, Yamadaoka, Suita, Osaka 565-0871, Japan}

\begin{abstract}
We present a possible theoretical interpretation of the observed afterglow emission of 
long gamma-ray burst GRB~080710.
While its prompt GRB emission properties are normal, 
the afterglow light curves in the optical and infrared bands are exceptional in 
two respects.
One is that the observed light curves of different wavelengths have 
maximum at the same time,
and that the achromatic peak time, $2.2\times10^3$~s after the burst trigger,
is about an order of magnitude later than typical events.
The other is that the observed flux before the peak increases more slowly than theoretically expected so far.
Assuming that the angular distribution of the outflow energy is top-hat or Gaussian-shaped,
we calculate the observed light curves of the synchrotron emission from the relativistic jets
and explore the model parameters that explain the observed data.
It is found that a narrowly collimated Gaussian-shaped jet with large isotropic-equivalent energy
is the most plausible model for reproducing the observed afterglow behavior.
Namely, an off-axis afterglow scenario to the achromatic peak is unlikely.
The inferred values of the opening angle and the isotropic equivalent energy of the jet 
are possibly similar to those of GRB~221009A, but the jet of GRB~080710 has a much smaller 
efficiency of the prompt gamma-ray emission.
Our results indicate a greater diversity of the GRB jet properties than previously thought.
\end{abstract}

\begin{keywords}
--- gamma-ray bursts: general
--- gamma-ray bursts: indivisual (GRB~080710)
\end{keywords}
 
\ExplSyntaxOn
\keys_set:nn { stm / mktitle } { nologo }
\ExplSyntaxOff
\maketitle


\section{Introduction}
Gamma-ray bursts (GRBs) are thought to originate in relativistically moving collimated outflows, details of which are still unknown 
\citep[e.g][]{piran1999,kumar2015,zhang2018}.
Prompt GRB emissions are followed by afterglows that tell us
various information on the relativistic jets.
The afterglows in the X-ray, optical/infrared, and radio bands
most likely arise from synchrotron radiation of
high-energy electrons
accelerated at relativistic shocks propagating into
circumburst medium \citep[e.g.][]{sari1998,sari1999}
and/or into the relativistic jets \citep[e.g.][]{sari1999b,kobayashi2000}.
Since {\it Neil Gehrels Swift Observatory} launched,
well-sampled early afterglow light curves in various wavelengths
have been collected, and it has been found that they 
contradict the 
expectations from
the standard afterglow model in the pre-{\it Swift} era.
The observed canonical X-ray afterglow unexpectedly consists of
three distinct phases \citep{nousek2006,zhang2006}.
In the optical/infrared bands, 
the pre-{\it Swift} afterglow model predicted
that light curves had
maxima at around $\sim10^{3-4}$~s after the prompt emission
\citep{sari1998},
and that peak times were later for lower frequencies.
Such chromatic behavior has not been firmly confirmed as of yet,
but instead, some bright events have shown 
the optical peak at $\sim10^2$~s from the beginning of the prompt emission,
which has been interpreted as the afterglow onset, that is,
the jet deceleration becomes significant at the peak time
\citep[e.g.][]{molinari2007,liang2010}. 
In the radio bands, it has been found that for  some events, 
like the recent bright burst GRB~221009A \citep{laskar2023,bright2023},
simple single-component models do not reproduce the observed light curves.
At present,
it may be important to study events showing 
observational features apparently inconsistent with pre-{\it Swift} afterglow theories.
Such theoretical works may provide us hints leading to new understandings of the nature of the GRB jet.

GRB~080710 was detected by {\it Swift} Burst Alert Telescope (BAT) \citep{GCN-7957},
and its duration was $T_{90}=120 ~\rm{s}$ \citep{GCN-7969}, so that this event is classified as a long GRB.
Its redshift is $z=0.845$ \citep{GCN-7962}.
The isotropic equivalent gamma-ray energy and
rest-frame peak energy in the $\nu F_\nu$-spectrum were estimated as
$E_{\gamma, \rm{iso}}=6\times10^{51}$~erg 
and $(1+z)E_{\rm peak}\simeq2\times10^2$~keV, respectively \citep{kruhler2009}.
These values are in the range of typical long GRBs observed so far
\citep[e.g.][]{minaev2020,zhao2020}.
The observed isotropic-equivalent luminosity
$L_{\gamma,{\rm iso}}=1-8\times10^{50}$~erg~s$^{-1}$ \citep{lu2012,liang2015,ghirlanda2018}
and $(1+z)E_{\rm peak}$ are also roughly consistent with Yonetoku relation
\citep[e.g.][]{yonetoku2004,li2023}.
On the other hand, GRB~080710 showed unusual
optical/infrared afterglow behavior \citep{kruhler2009}.
The light curves in various bands have maximum at the same time at $T=T_{\rm peak}\simeq2.2 \times 10^3$~s, 
where $T$ is the observer time after the BAT trigger.
Then, the r-band isotropic peak luminosity, $L_\nu\sim 1.6 \times10^{31}$~erg~s$^{-1}$~Hz$^{-1}$,
is in the same order as other events with optical afterglow detection 
 \citep[e.g.,][]{naradini2008,nysewander2009,cenko2009,kann2010,panaitescu2011,liang2013,panaitescu2013}.
Before the peak, the observed flux arises with $F_\nu\propto T^{1.1}$.
According to the pre-{\it Swift} standard model,
the optical peaks should have been chromatic when they occur at $T\sim10^{3-4}$~s,
and the flux before the peak time should increase more gradually 
than observed
\citep[$F_\nu\propto T^{1/2}$:][]{sari1998}.
Although an interpretation of afterglow onset might explain the observed achromatic behavior,
the observed peak time is an order of magnitude later than usual
\citep{panaitescu2011}, and then the rising part
would have been more steeper than observed
\citep[$F_\nu\propto T^{2-3}$:][]{sari1999b}.
\citet{kruhler2009} proposed that the observed achromatic peak
is a signature of off-axis afterglow 
\citep[e.g.,][]{granot2002,vaneerten2010}.
However,
also in this case, 
the flux before the peak would have increased more rapidly than observed.
Moreover, the prompt emission should have been
dimmer and softer than observed
\citep{ioka2001,ioka2018,ioka2019,ramirez-ruiz2005,
sato2021,salafia2015,salafia2016,yamazaki2002,yamazaki2003a,yamazaki2004,yamazaki2003b}.
So far, similar events with late-time, achromatic optical peaks
as GRB~080710 have been detected, such as
GRBs~050408 \citep{postigo2007},
071031 \citep{kruhler2009-XRF}, 
and 080603A \citep{guidorzi2011}.

In this paper, 
using
Bayesian inference with the aid of Markov chain Monte Carlo (MCMC) sampling,
we discuss on the possible explanation of
achromatic peak of GRB~080710 afterglow. 
In the previous paragraph, we did not consider the angular structure of the jet,
in other words, the issues on this event were based on the uniform, top-hat (TH) jet.
They might be reconciled by structured jets 
\citep[e.g.,][]{rossi2002,zhang2002_p,zhang2004_g,kumar2003,beniamini2020a,oganesyan2020}.
When the GRB jet has angular dependence of the initial kinetic energy and/or bulk Lorentz factor
(that is, they are functions of the angle from the central axis of the jet),
the observed afterglow 
light curves are different from those for the TH case.
The jet structure contains rich information on the jet formation and/or propagation
into dense media around the central engine
\citep[e.g.,][]{zhangW2003,morsony2007,mizuta2013,gottlieb2021}.
It had long been desired to be studied until the detection of a nearby short GRB~170817A with precise
multiwavelength data.
A lot of  authors take this opportunity to reconsider the jet structure, and it has been common
that the TH jet model is difficult to explain the observation
\citep[e.g.,][]{davanzo2018,lazzati2018,lyman2018,margutti2018,troja2018,xie2018,ghirlanda2019,ioka2019,lamb2019,
beniamini2020b,takahashi2020,takahashi2021,mooley2022}.
Such argument has also been applied to long GRBs
160625 \citep{cunningham2020}, 221009A 
\citep{connor2023,sato2023,gill2023,lhaaso2023}, and so on.

This paper is organized as follows.
In Section~\ref{section-title:modeling}, 
we introduce models of afterglow emission from the structured jets
to explain the observed data of GRB~080710, and describe our fitting method.
In Section~\ref{section-title:result}, 
we show the results for the models, and discuss their differences among them.
Section~\ref{section_discussion} is devoted for discussion.
Following \citet{kruhler2009}, we adopt the flat $\Lambda$CDM model with
cosmological parameters,
$H_0=71$~km~s$^{-1}$, 
$\Omega_M =0.27$, and 
$\Omega_{\Lambda} =0.73$ \citep{spergel2003}.
Then, the luminosity distance is calculated as
$d_L=1.67\times10^{28} ~{\rm cm}$ 
for the redshift $z=0.845$.

\section{Afterglow Model and Fitting Method}
\label{section-title:modeling}

In this section, we present afterglow emission models and
a method of parameter estimation using MCMC.
We use an open-source Python package \texttt{afterglowpy} \citep{Ryan2020}
that numerically calculates the synchrotron emission from 
a shell relativistically moving  into uniform ambient medium with
a number density $n_0$.
Let $(r,\theta)$ be radial and polar angle coordinates in the
rest frame of the central engine, located at $r=0$.
The central axis of the axisymmetric jet corresponds to $\theta=0$.
In this frame, we define
an initial
angular distribution of the isotropic equivalent energy of the shell,
$E(\theta)=4 \pi (dE/ d\Omega)$.
In this paper, we adopt two cases for $E(\theta)$: TH and Gaussian-shaped jets.
The former is expressed as
\begin{align}\label{tophatjet}
    E(\theta) = \left\{ \begin{array}{ll}
    E_0 & (\theta \leqq \theta_j) \\
    0 & (\theta > \theta_j)
      \end{array} \right.~~,
\end{align}
where $E_0$ is constant, and $\theta_j$ is an opening half-angle.
On the other hand, the Gaussian jet model is described by
\begin{align}\label{gauusianjet}
    E(\theta) = \left\{ \begin{array}{ll}
    E_0 \exp{\left( -\frac{\theta^2}{2 \theta_c^2} \right)} & (\theta \leqq \theta_j) \\
    0 & (\theta > \theta_j)
      \end{array} \right.~~,
\end{align}
where the most of the jet energy is confined in $\theta<\theta_c$.
Note that the collimation-corrected energy of the jet is calculated as
\begin{align}\label{eq:collimation-corrected-energy}
E_{\rm K, jet} = 2 \int_{0}^{2 \pi} d\phi \int_{0}^{\theta_j}d\theta \, \sin{\theta} \frac{E(\theta)}{4 \pi}~~.
\end{align}
This integration can be approximated as
$E_{\rm K, jet}\approx E_0\theta_j^2/2$ and 
$E_{\rm K, jet}\approx E_0\theta_c^2[1-\exp(-\theta_j^2/2\theta_c^2)]$ 
for the TH and Gaussian jets, respectively,
when $\theta_j \ll 1$.
The angular distribution of the initial Lorentz factor of the shell, $\Gamma(\theta)$,
is given in two limiting cases.
One is the case of $\Gamma(\theta)= \Gamma_0$, where $\Gamma_0$ is constant,
so that initial Lorentz factor is uniform in $\theta$ (Gamma-Flat: hereafter, GF).
The other case is given by
$\Gamma (\theta) = 1 + (\Gamma_0 - 1)E(\theta)/E_0$, so that
the initial mass-loading is uniform in $\theta$ (Gamma-Even-Mass: hereafter, GEM).
Then, the evolution of the bulk Lorentz factor, $\Gamma(r,\theta)$, of the emitting material 
moving in the direction of $\theta$ is determined from the free expansion to
adiabatic deceleration phases. 
Once the jet dynamics is given, the observed synchrotron emission is calculated.
The index of the electron energy distribution, $p$,
and microphysics parameters,
$\epsilon_e$ (the fraction of the shock internal energy that is partitioned to electron),
$\epsilon_B$ (the fraction of the shock internal energy that is partitioned to magnetic fields),
and 
$\xi_N$ (the number fraction of electrons that are accelerated to a non-thermal distribution)
are assumed to be constant.
Observer's line of sight is in the direction $\theta=\theta_{\rm obs}$.

Using the Python MCMC module \texttt{emcee} \citep{MCMC},
we perform a Bayesian estimation to determine the best-fit
parameters explaining the observed data of 
the optical, infrared, and X-ray afterglow of GRB~080710.
For the TH jet model,
we will have posterior distributions for nine quantities,
$E_0$, $\Gamma_0$, $\theta_j$, 
$\beta$, $n_0$, $p$, $\epsilon_e$, $\epsilon_e$, and $\xi_N$ (Table~\ref{table:best_paramTH}),
while for the GF and GEM models,
another quantity $\alpha$ is added
(Tables~\ref{table:best_paramGF} and \ref{table:best_paramGEM}),
where we define
$\alpha = \theta_j / \theta_c$ and
$\beta = \theta_{\rm obs}/\theta_j$.
We adopt log-uniform prior distributions for $E_0$, $n_0$, $\epsilon_e$, $\epsilon_B$, and $\xi_N$,
and uniform prior distributions for $\Gamma_0$, $\theta_j$, $\alpha$, $\beta$, and $p$.
We impose the condition $\alpha\geq1$ (i.e., $\theta_j \geq \theta_c$) to have 
Gaussian-like structure.
A general case, $0\le \beta \le 1.5$, and
an off-axis viewing case, $1\le \beta \le 1.5$,
as prior ranges are studied in the next section.
We set $2.01\le p\le 2.9$ as a prior range because \texttt{afterglowpy} works for $p>2$.
The other parameters have sufficiently wide prior range.
The likelihood for each nine- or ten-parameter set  is calculated by 
\begin{align}\label{likelihood_func}
  {\cal L}(\vec{\theta}) 
  = \prod_{k=1}^N \frac{1}{\sqrt{2\pi \sigma_k^2}} \exp \left({-\frac{(Y_k-F_k(\vec{\theta}))^2}{2\sigma_k^2}}\right),
\end{align}
where $N$ is the number of observation data points, 
$Y_k$ and $\sigma_k$ are the median  and the standard deviation (1$\sigma$ error) 
of the $k$-th observed flux ($k=1,2,\cdots,N$), respectively,
and
 $\vec{\theta}$ represents model parameters which are
 necessary to theoretically calculate the flux $F_k$.
We generate 32 independent MCMC chains with $10^5$ steps to get the posterior distributions. 
In order to get the sufficiently converged samples,
we discard the first $\sim2\times10^4$ steps as burn-in for TH and GF cases, while the first $\sim1\times10^3$ steps for GEM case. 
Samples once per 100 steps for each MCMC chain are obtained.

\section{Results}
\label{section-title:result}

In this section, the results of our parameter fitting analysis, 
are shown for the three cases (TH, GF, and GEM).
For each case, we investigate the behavior of the best-fit model for GRB~080710 afterglow in 
z-band ($\nu_z=3.3\times10^{14}$~Hz), r-band ($\nu_r=4.8\times 10^{14}$~Hz), 
and X-ray band ($\nu_X=7.3\times 10^{16}$~Hz).

The X-ray data are extracted from the
\textit{Swift} team website\footnote{https://www.swift.ac.uk/xrt\_curves/00316534/} \citep{evans2007,evans2009}.
The data provides us with time evolution of the integrated energy flux in the 0.3--10~keV and spectral hardness. 
The spectral hardness is almost constant with time, and
the time-averaged photon index is $1.92^{+0.12}_{-0.11}$. 
Hence, we assume the photon index of 1.92 all the time to calculate the energy flux density at 
$h\nu_X=0.3$~keV.
In the MCMC fitting, we adopted a uniform error of 1\% for all X-ray observed data points to ensure equal weighting of the data compared to the optical/infrared ones.
The r-band and z-band flux densities were taken from \citet{kruhler2009}.
Here, we thin out observed r-band and z-band data
to get roughly equal number of data points in a logarithmic time intervals,
reducing the weight of dense data from
large number of observation points after the peak time.
%
%
We add systematic error of 0.012~mag for both bands 
\citep[see online material\footnote{Strasbourg astronomical Data Center (CDS):\\
http://cdsarc.u-strasbg.fr/viz-bin/qcat?J/A+A/508/593\#/prov} for][]{kruhler2009}.
The values of extinction, $A_{\rm r}$ and $A_{\rm z}$, are calculated as follows.
The extinctions in r-band (0.63~$\mu$m) and z-band (0.92~$\mu$m)  in our Galaxy toward the direction of GRB~080710
are estimated as
$A_{\rm r}^{\rm MW}=0.171~{\rm mag}$ and $A_{\rm z}^{\rm MW}=0.094~{\rm mag}$, respectively \citep{schlafly2011}.
Wavelengths of these lights are $\lambda=0.34~\mu$m and 0.50~$\mu$m at the rest frame of the host galaxy.
The V-band (0.55~$\mu$m) extinction is measured as $A_{\rm V}^{\rm HG} = 0.03 \pm 0.01$ \citep{PSchady}.
Assuming the formula of extinction law given by \citet{Cardelli+89},
we convert $A_{\rm V}^{\rm HG}$ into $A_\lambda^{\rm HG}$ of the
other bands, and get
$A_{\rm r}^{\rm HG} = 0.0485 ~{\rm mag}$ and $A_{\rm z}^{\rm HG} = 0.0337 ~{\rm mag}$, respectively.
Hence, the total extinctions are derived as
$A_{\rm r} = A_{\rm r}^{\rm HG}+A_{\rm r}^{\rm MW}=0.2195~{\rm mag}$
and
$A_{\rm z} = A_{\rm z}^{\rm HG}+A_{\rm z}^{\rm MW}=0.1277 ~{\rm mag}$.

Median values for 
the one-dimensional projection of the (marginalized) posterior distributions of each model parameter,
as well as the best-fit parameters,
for TH, GF, and GEM models are shown in 
Tables~\ref{table:best_paramTH},~\ref{table:best_paramGF} and~\ref{table:best_paramGEM}, respectively,
where we set prior range $0\leq\beta\leq1.5$.
It is found that
for all three cases,
the median and the best-fit values are close with each other.

Table~\ref{table:best_param_sum} summarises
the best-fit parameters derived for the cases of the 
different prior ranges of $\beta=\theta_{\rm obs}/\theta_j$
such that $0\leq\beta\leq1.5$ and $1\leq\beta\leq1.5$.
The limitation to the off-axis viewing cases (that is, the
narrower prior range of $\beta$) does not produce
significantly better fit for the three models (TH, GF, and GEM).
Furthermore, in the 
case of the
prior range $1\leq\beta\leq1.5$,
the best fitted values of $\beta$ is very close to unity,
so that we cannot see the characteristics of the off-axis afterglow.
Hence, in the following sections~\ref{subsection:TH}, \ref{subsection:GF}, and \ref{subsection:GEM}, we explain the physical behavior of the best-fit model obtained for the general prior range $0 \leq \beta \leq 1.5$.

\begin{figure*}[ht]
\centering 
\includegraphics[width=\linewidth]{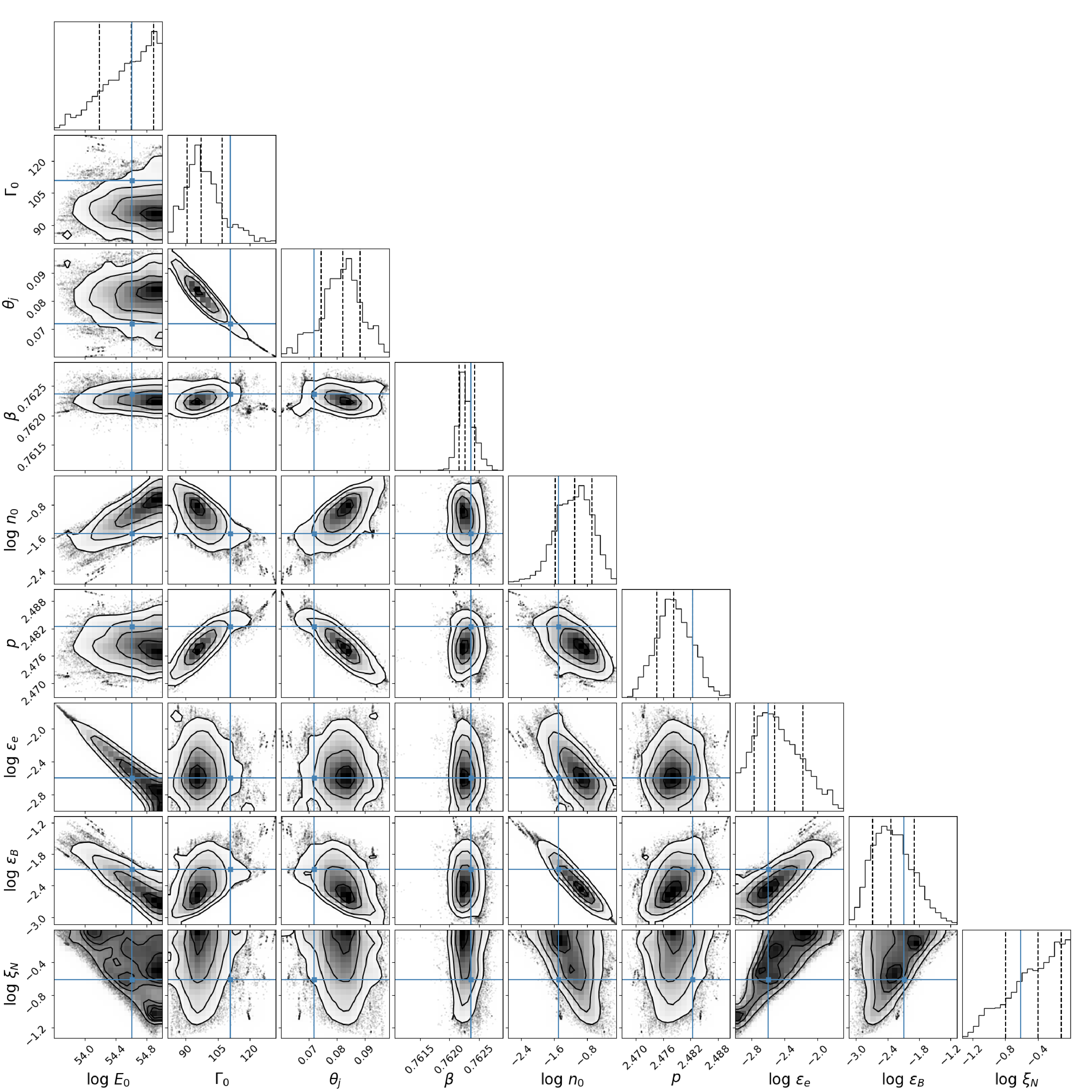}
\caption{
Posterior probability distributions of parameters by TH jet model for GRB~080710 afterglow.
We assume flat prior with a range $0\leq\beta\leq1.5$ in the MCMC sampling.
Contours of 0.5, 1, 1.5, and 2$\sigma$ are shown in the two-dimensional space of all possible combinations of two model parameters.
Vertical dashed lines represent the median value and  68\% uncertainties (the 16\% and 84\% quantiles) for each parameter.
Blue solid lines show the location of the best-fit parameters (giving the maximum posterior probability).
}
\label{figure:psterior-TH}
\end{figure*}

\begin{figure*}[ht]
\centering 
\includegraphics[width=\linewidth]{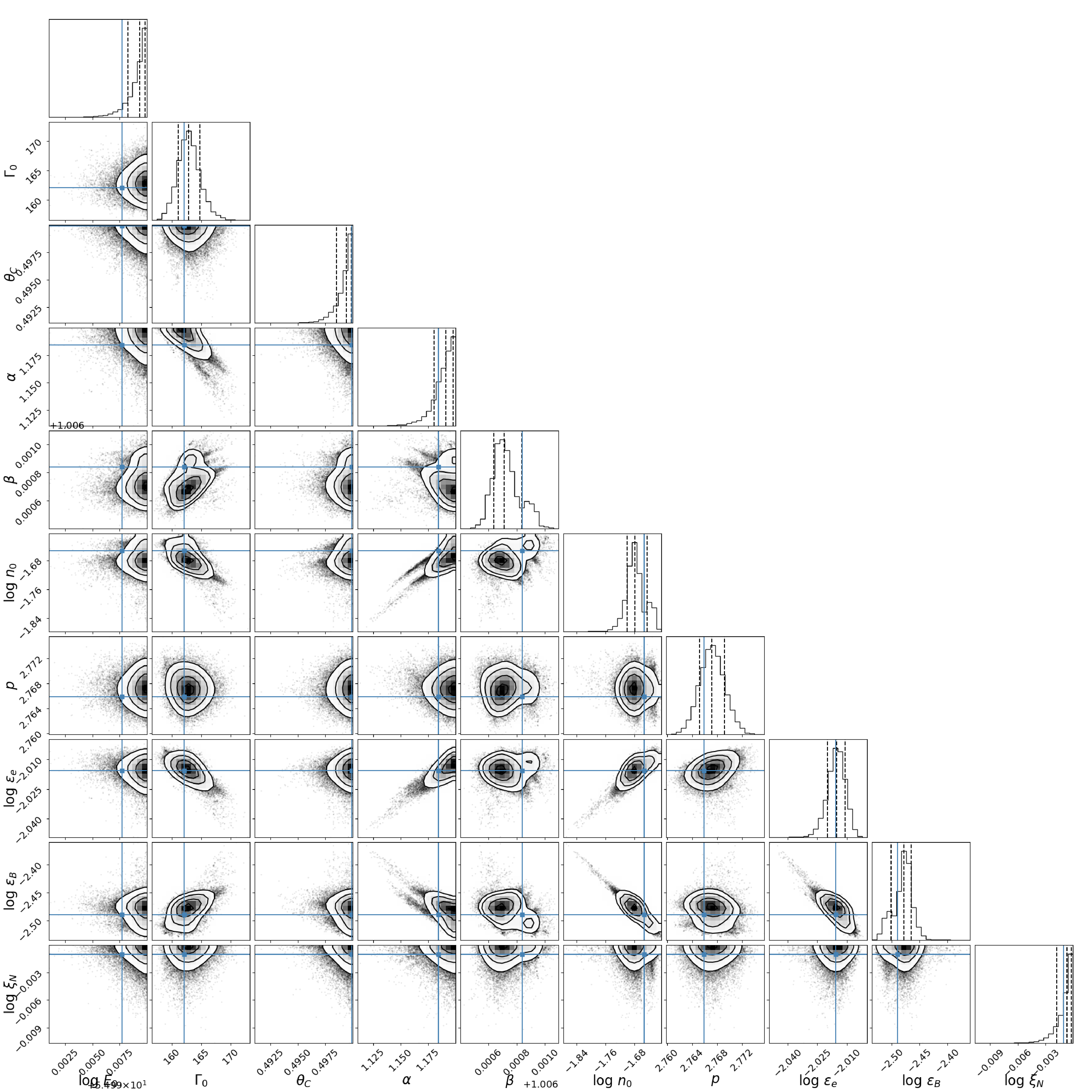}
\caption{ 
The same as Fig.~\ref{figure:psterior-TH} but for the GF model.
}
\label{figure:psterior-GF}
\end{figure*}

In Fig.~\ref{lc-all}, we draw the z-band (blue), r-band (red), and X-ray (green) 
light curves for the best-fit parameters in the cases of
TH (dashed lines), GF (dotted lines), and GEM (solid lines).
Figure~\ref{lc-aroundAP} is an enlarged view of Fig.~\ref{lc-all},
showing the differences among the three models around the peak time.
In particular,
the rising slope predicted by the GEM is the most gradual.
We show in Fig.~\ref{AP} 
the ratio of r-band ($F_r$) to z-band ($F_z$) fluxes.
The covariances and posterior probability distributions of model parameters obtained by
MCMC sampling for TH, GF, and GEM models are shown in 
Figs.~\ref{figure:psterior-TH}, \ref{figure:psterior-GF}, and \ref{figure:psterior-GEM}, respectively.
As shown in Table~\ref{table:best_param_sum}, the GEM model with prior range $0\leq\beta\leq1.5$ gives
the smallest value of $\chi^2$/{\it d.o.f.}.

In the following Sections~\ref{subsection:TH}, \ref{subsection:GF}, and \ref{subsection:GEM},
we describe the properties of the best-fit parameter sets for TH, GF, and GEM models, respectively.
For this purpose, we introduce several timescales and characteristic frequencies,
which are analytically estimated in Appendix.
We define
$T_{\rm dec}(\theta)$ as the time 
at which
 photons  emitted from the  region  moving into $\theta$-direction 
that starts to decelerate arrive at the observer,
$T_{\rm beam}(\theta)$ as the time 
at which photons emitted from the  region  moving into $\theta$  
whose bulk Lorentz factor satisfies $\Gamma(r,\theta)=|\theta_{\rm obs}-\theta|^{-1}$ arrive at the observer,
and
$T_{\rm jet}$ as the jet break time, that is,
the time when the whole jet becomes visible to the observer.
Note that $T_{\rm beam}(\theta)$ is defined in the case
where the emission region 
satisfies
an initial condition 
$\Gamma(\theta)>|\theta_{\rm obs}-\theta|^{-1}$, and
the observed flux from the emitting matter with $\theta$
is dim until $T_{\rm beam}(\theta)$ due to the relativistic beaming
effect.
Furthermore, the emission region with $\theta$ 
provides characteristic frequencies, $\nu_m(T, \theta)$ and $\nu_c(T, \theta)$, 
at the observer time $T$, 
of synchrotron emission from electrons with 
the minimum Lorentz factor 
and the cooling Lorentz factor, respectively, and their crossing times $T_m(\theta,\nu)$ and $T_c(\theta,\nu)$
of the observing frequency, $\nu$,
respectively.

\subsection{The case of Top-hat (TH) jet}
\label{subsection:TH}

\begin{figure*}[ht]
\centering 
\includegraphics[width=\linewidth]{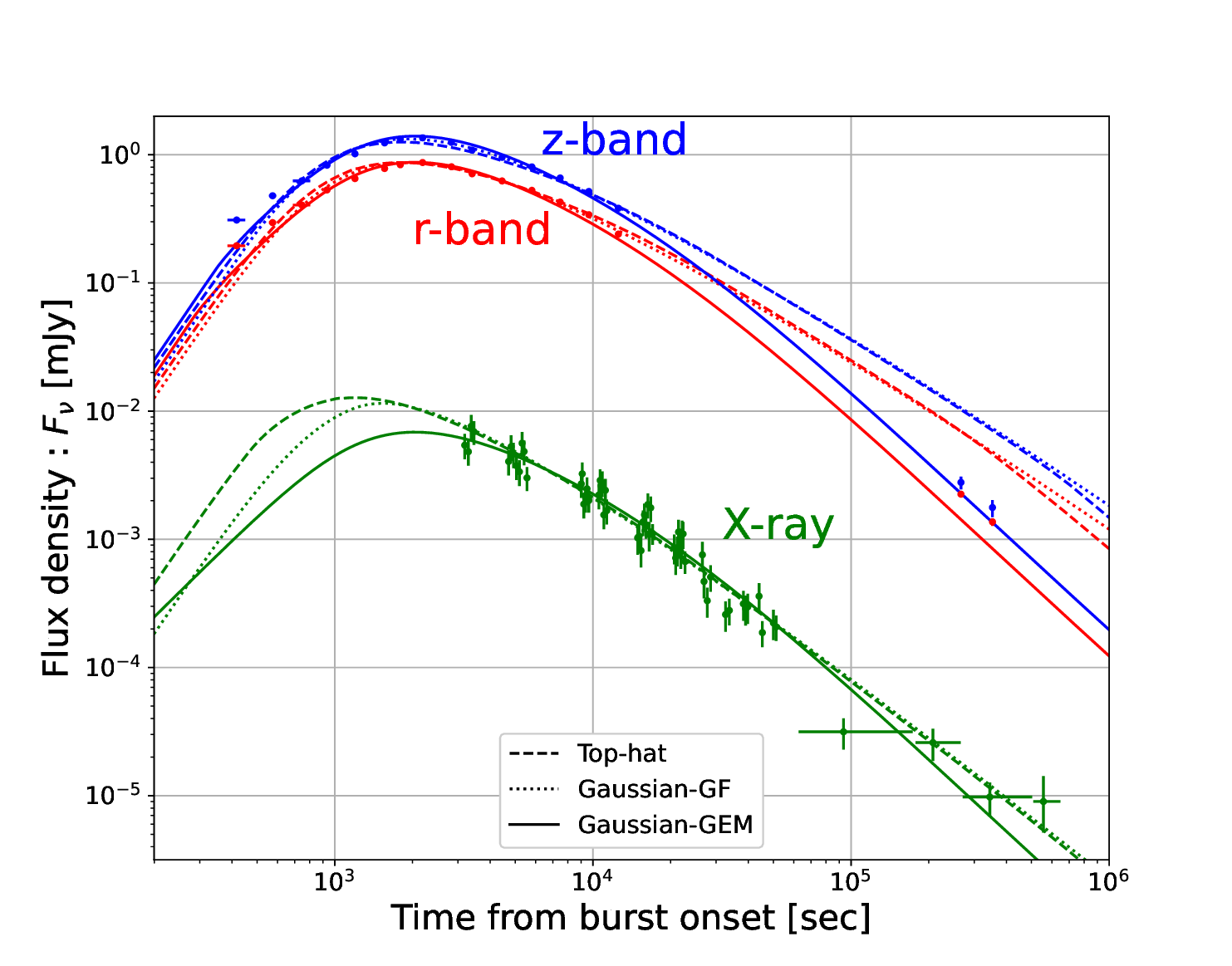}
\caption{
The z-band (blue), r-band (red), and X-ray (green) light curves of GRB~080710.
Observed data are compared with 
theoretical results with the best-fit parameter sets of Top-hat (TH: dashed lines), Gaussian-GF (GF: dotted lines) 
and Gaussian-GEM (GEM: solid lines) jet models, where we impose a prior range $0\leq\beta\leq1.5$ in the MCMC sampling.
The best-fit parameters are
$E_0 = 4.09 \times 10^{54} ~{\rm erg}$, 
$\Gamma_0 = 1.11\times 10^2$, 
$\theta_j = 7.19\times 10^{-2}~{\rm rad}$, 
$\theta_{\rm obs} = 5.48\times 10^{-2}~{\rm rad}$, 
$n_0 = 3.24\times 10^{-2} ~{\rm cm^{-3}}$, 
$p=2.48$, 
$\epsilon_e = 2.50\times 10^{-3}$, 
$\epsilon_B = 8.13\times 10^{-3}$, 
and
$\xi_N = 0.246$
for the TH,
$E_0 = 9.95 \times 10^{54} ~{\rm erg}$, 
$\Gamma_0 = 1.62\times 10^2$, 
$\theta_c = 5.00\times 10^{-1} ~{\rm rad}$, 
$\theta_j = 5.92\times 10^{-1} ~{\rm rad}$, 
$\theta_{\rm obs} = 5.96\times 10^{-1} ~{\rm rad}$, 
$n_0 = 2.23\times 10 ~{\rm cm^{-3}}$, 
$p=2.77$, 
$\epsilon_e = 9.65\times 10^{-3}$, 
$\epsilon_B = 3.24\times 10^{-3}$, 
and
$\xi_N = 0.997$,
for GF and
$E_0 = 9.98 \times 10^{54} ~{\rm erg}$, 
$\Gamma_0 = 1.04\times 10^2$, 
$\theta_c = 1.21\times 10^{-2} ~{\rm rad}$, 
$\theta_j = 2.42\times 10^{-2} ~{\rm rad}$, 
$\theta_{\rm obs} = 1.15 \times 10^{-4}~{\rm rad}$, 
$n_0 = 3.47\times 10^{-2} ~{\rm cm^{-3}}$, 
$p=2.01$, 
$\epsilon_e = 1.12\times 10^{-2}$, 
$\epsilon_B = 4.99\times 10^{-1}$, 
and
$\xi_N = 0.543$
for GEM 
(see Tables~\ref{table:best_paramTH}, 
\ref{table:best_paramGF}, \ref{table:best_paramGEM}, and the left 3 columns in Table~\ref{table:best_param_sum}).}
\label{lc-all}
\end{figure*}

\begin{figure}[ht]
\centering 
\includegraphics[width=\linewidth]{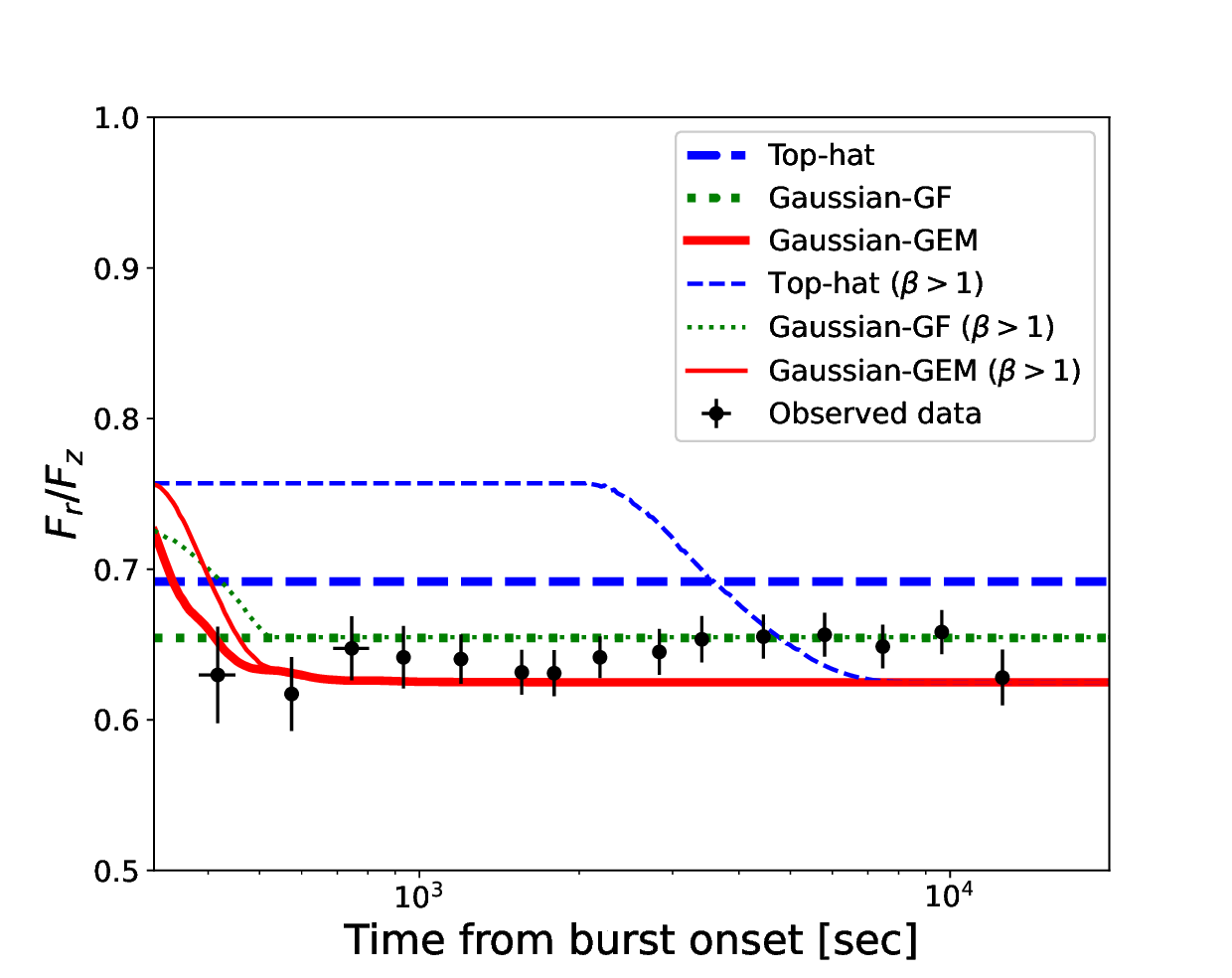}
\caption{
The flux ratio of r-band to z-band afterglows, $F_r/F_z$, of GRB~080710.
Observed data (black points) are compared with 6 theoretical results with the best-fit parameter sets with 
TH (blue-dashed line), GF (green-dotted line), GEM (red-solid line) models
shown in Table~\ref{table:best_param_sum}.
Thick and thin lines are for cases of different 
prior ranges $0\leq\beta \leq 1.5$ and $1\leq\beta \leq 1.5$ in the MCMC fitting,
respectively.}
\label{AP}
\end{figure}

Dashed lines in Fig.~\ref{lc-all} 
represent the best-fit TH model obtained in
the case of prior range $0\leq\beta\leq1.5$.
The fit is not acceptable ($\chi^2/{\it d.o.f}=22.5$  as shown in Table~\ref{table:best_param_sum}).
The posterior distributions of the model parameters are shown in
Fig.~\ref{figure:psterior-TH}.
In particular, the posterior distribution of $\beta$ peaks around 0.76,
so that the observer's line of sight is inside the jet cone ($\theta_{\rm obs}<\theta_j$).
Nevertheless, we perform the MCMC fitting imposing the prior range
$1\leq\beta\leq1.5$,
in order to see if the observed achromatic peak is a signature of
off-axis afterglow as claimed by \citet{kruhler2009}.
Although the fitting result becomes slightly better
($\chi^2/{\it d.o.f.}=18.3$ as shown in Table~\ref{table:best_param_sum}),
the fit is still unacceptable, and
the rising part before the peak becomes steeper, and the
achromaticity around the peak disappeared (see blue thin dashed line in Fig.~\ref{AP}).
Furthermore, the best fitted value of $\beta$ is only slightly larger than unity ($\beta-1=1.5\times10^{-3}$),
so that the off-axis afterglow rising behavior cannot be seen.
Hence, below in this section 3.1, 
we explain the physical behavior of the best-fit model obtained
for the general prior range $0\leq\beta\leq1.5$.

\begin{figure}[ht]
\centering 
\includegraphics[width=\linewidth]{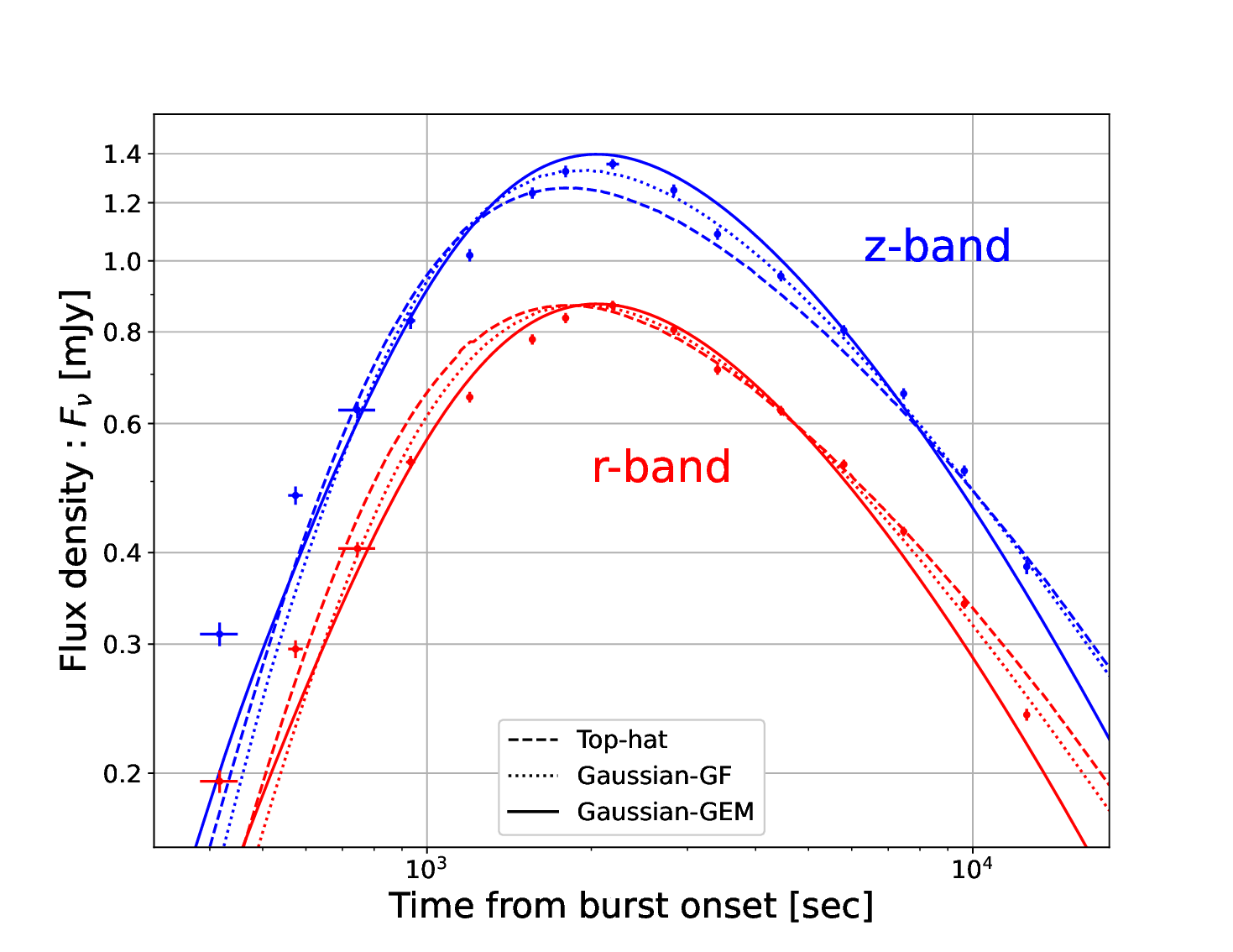}
\caption{
Enlarged view of Fig.~\ref{lc-all} around peak time ($T\sim 2\times10^3 ~{\rm s}$) in
the r-band (red) and z-band (blue).
}
\label{lc-aroundAP}
\end{figure}

Figures~\ref{lc-all}--\ref{AP} show the achromatic peaks at $\sim 2\times10^3$~s
in the z and r-bands, though the TH model (dashed-lines) overpredicts the observed flux at $T>10^{4.5} ~{\rm s}$.
Moreover, as can be seen in Fig.~\ref{lc-aroundAP}, the rising part before the peak
is steeper than observed.
To understand the behavior of the light curves, let us consider the emission from the region toward us
($\theta \approx \theta_{\rm obs}$).
In the TH case, if $\theta_{\rm obs} <\theta_j$, the shape of the light curves hardly depends on $\theta_{\rm obs}$.
For our inferred parameters, the deceleration time is calculated as 
$T_{\rm dec}(\theta_{\rm obs}) \sim 10^{3} ~{\rm s}$
[see Eq.~(\ref{eq:Tdec})],
which is near the observed peak.
We also find $\nu_m(T, \theta_{\rm obs})<\nu_z<\nu_r<\nu_c(T, \theta_{\rm obs})$ around
$T=T_{\rm dec}(\theta_{\rm obs})$.
Then, the r-band and z-band fluxes behave as
$F_{\nu} \propto T^3$ for $T<T_{\rm dec}(\theta_{\rm obs})$ \citep{sari1999b},
and $F_{\nu} \propto T^{-3(p-1)/4}=T^{-1.11}$ for $T>T_{\rm dec}(\theta_{\rm obs})$ 
and $p=2.48$ \citep{sari1998}.
Therefore, the r-band and z-band light curves take maximum simultaneously at $T\sim T_{\rm dec}(\theta_{\rm obs})$ 
due to the transition from the free expansion to the deceleration phases.

The deceleration time 
$T_{\rm dec}(\theta_{\rm obs}) \sim10^{3} ~{\rm s}$
requires $E_0=4\times10^{54}$~erg
that is somewhat larger than the value of typical long GRBs.
Since $T_{\rm dec}(\theta_{\rm obs}) \propto (E_0/n_0)^{1/3}\Gamma_0^{-8/3}$
[see Eq.~(\ref{eq:Tdec})], we might have small $n_0$ and/or $\Gamma_0$ 
keeping smaller $E_0$ ($\sim10^{52-53}$~erg).
However, in this case, the peak flux at $T_{\rm dec}(\theta_{\rm obs})$, which is calculated as
$F_{\rm peak} \propto \xi_N^{-0.48} \epsilon_e^{1.48} \epsilon_B^{0.84} n_0^{0.84}\Gamma_0^{2.96} E_0$
for $p=2.48$, 
should have been smaller than observed.
Once $E_0$ is large, small value of $\epsilon_e=2.5\times10^{-3}$ is necessary for
the observed peak flux of $\sim 1$~mJy.

The cooling break in the observed spectrum by emitting matter moving toward 
the observer, $\nu_c(T, \theta_{\rm obs})$,
crosses the X-ray band at 
$T_c(\theta_{\rm obs}, \nu_X)\sim 7 \times 10^{2} ~{\rm s}$
[see Eq.~(\ref{eq:time_nu_c_on-free})], after which
$\nu_c(T, \theta_{\rm obs})<\nu_X$.
Since inferred electron spectral index is $p=2.48$,
the X-ray photon index should have been $(p/2)+1=2.24$
 in the epoch of X-ray observation ($T\gtrsim3\times10^3$~s).
This may be inconsistent with the observed value of $1.92_{-0.11}^{+0.12}$.
It is also found that around $T_c(\theta_{\rm obs}, \nu_X)\sim 7 \times 10^2$~s,
rising slope of the X-ray light curve 
changes to be shallower,
leading to the slightly earlier maximum
compared with r and z-bands (see Fig.~\ref{lc-all}).

As shown in Fig.~\ref{lc-all},
the power-law decay after the peak time $T_{\rm peak}$ lasts until $T\sim6\times10^5$~s, which
is followed by more rapid decay.
This steepening can be interpreted to be the jet break, that is, the transition to the jetted 
afterglow phase, because
we estimate the jet break time as
$T_{\rm jet} \sim 10^6$~s 
[see Eq.~(\ref{eq:jet_break})].

\subsection{The case of Gaussian jet 1: Gamma-Flat (GF)}
\label{subsection:GF}

The best-fit GF model obtained by MCMC sampling with the prior range
$0\leq\beta\leq1.5$ (Table~\ref{table:best_paramGF})
provides the afterglow light curves shown in dotted lines in Fig.~\ref{lc-all}.
In this case, 
the value of
reduced chi-squared is 24.7, so that the fit is not acceptable.
The posterior distributions of the model parameters are shown in 
Fig.~\ref{figure:psterior-GF}.
The MCMC fitting for the GF model infers large $E_0$ ($\sim10^{55}$~erg), $\theta_j$ ($\sim0.5$~rad), and $\xi_N$ ($\sim1$),
all of which are near the upper end of each prior range.
Since we adopt sufficiently wide prior ranges on these parameters 
taking into account the values of other long GRBs 
\citep[e.g.,][]{zhao2020},
the GF model is already inadequate for an explanation of the observed 
data.
As seen in the TH case 
(see Table~\ref{table:best_param_sum} and \S~\ref{subsection:TH}), 
it might be possible that 
if we take narrower prior range for $\beta$ (that is, $1.0<\beta<1.5$),
the fit would become better.
It is also to be studied for structured jets
if the off-axis viewing case for the achromatic peak is viable 
as proposed by \citet{kruhler2009}.
However, the MCMC result for the case of $1.0<\beta<1.5$
is almost unchanged (see Table~\ref{table:best_param_sum}).
Hence, in the following of this subsection, we briefly present
the physical explanation of the model with the best-fit parameter set
obtained for the prior range $0\leq\beta\leq1.5$.

Figures~\ref{lc-all} and \ref{lc-aroundAP} show that
the model provides optical/infrared achromatic peak at $\sim10^3$~s.
The r-band and z-band fluxes roughly obeys the scaling, $F_\nu\propto T^3$, before the peak.
Since the initial jet energy $E(\theta)$ has angular dependence,
the quantitative discussion is limited. However,
the behavior of the best-fit model can be qualitatively understood as in the following.
%
%
Using the inferred values listed in Table~\ref{table:best_paramGF},
we find 
initially  $\theta_{\rm obs} > \theta_j$ and
$\Gamma_0(\theta_{\rm obs}-\theta_j) = 0.66 (< 1)$, so that
the emission from the jet edge ($\theta=\theta_j=0.592$~rad) can be seen from
 the observer before the beginning of the jet deceleration.
Since $\Gamma(\theta) = \Gamma_0$ for the GF model (Section~\ref{section-title:modeling}),
we approximately obtain $T_{\rm dec}(\theta)\propto \exp{(-\theta^2/6\theta_c^2)}$
for the emission region moving in the direction $\theta$ whose range is
$0.590~{\rm rad}\approx (\theta_{\rm obs}-\Gamma_0^{-1}) < \theta \leq \theta_j$ [see Eq.~(\ref{eq:Tdec})],
and we find
$T_{\rm dec}(\theta_j)\sim T_{\rm dec}(\theta_{\rm obs}-\Gamma_0^{-1})\sim10^3$~s,
which is comparable to the observed peak time $T_{\rm peak}$.
The emission from the region with $\theta<\theta_{\rm obs}-\Gamma_0^{-1}$
is very dim and negligible until 
$T\sim T_{\rm beam}(\theta_{\rm obs}-\Gamma_0^{-1})$.
Since $T_{\rm beam}(\theta_{\rm obs}-\Gamma_0^{-1})\sim10^3$~s,
the observed flux around $T_{\rm peak}$ is dominated by the emission region with
$0.590~{\rm rad} \lesssim \theta \leq \theta_j$.
At early time the jet is in the free expansion phase, and
we numerically find
$\nu_m(T, \theta) < \nu_z < \nu_r  <  \nu_X < \nu_c(T, \theta)$, 
resulting in the scaling for the rising part, $F_{\nu} \propto T^3$, for $T<T_{\rm dec}(\theta)$.
In particular, the numerically calculated light curves shown in Fig.~\ref{lc-all}
can be well explained by this scaling until $T\sim6\times10^2$~s.
Between $T\sim6\times10^2$~s and the peak time $T_{\rm peak}$, 
the total flux is partially
contributed by the decelerating region, which has
$T_m(\theta,\nu_z)<T_{\rm dec}(\theta)$ and
$\nu_m(T, \theta) < \nu_z < \nu_r  <  \nu_c(T, \theta)<\nu_X$.
Hence, the optical/infrared peak is achromatic.
After the peak, the flux scales as
$F_{\nu} \propto T^{-3(p-1)/4} = T^{-1.3}$ with $p = 2.77$,
resulting in slow decay, overshooting 
observed r-band and z-band data taken in the late epoch ($T\sim3\times10^5$~s).
The X-ray photon index is predicted as $p/2+1=2.39$,
which is inconsistent with the observed value of $1.92_{-0.11}^{+0.12}$.
Similarly to the TH case, the large value of $E_0 = 1 \times 10^{55}$~erg  and
the small value of $n_0 = 2 \times 10^{-2} ~{\rm cm^{-2}}$ as well as
 $\epsilon_e = 0.01$ are necessary for
$T_{\rm dec}(\theta_j) \sim T_{\rm peak} \sim 10^3$~s
and the observed peak flux of $\sim {\rm mJy}$.

The emission from the region moving into the direction $\theta$ that 
is much smaller than $\sim(\theta_{\rm obs}-\Gamma_0^{-1})=0.590$~rad 
(that is, $\Gamma_0(\theta_{\rm obs}-\theta)>1$)
is negligible to the observed flux around $T_{\rm peak}$.
We find, from Eqs.~(\ref{eq:Tdec}) and (\ref{eq:beam_time}),
$T_{\rm dec}(\theta) \propto (\theta_{\rm obs}-\theta)^2\exp{(-\theta^2/6\theta_c^2)}$,
which increases with decreasing $\theta$,
and the ratio
$T_{\rm beam}(\theta)/T_{\rm dec}(\theta) = [\Gamma_0(\theta_{\rm obs} - \theta)]^{2/3}>1$.
The flux that is received by the observer from the region with $\theta$ takes maximum at $T_{\rm beam}(\theta)$ 
if $T_{\rm dec}(\theta)<T_{\rm beam}(\theta)$.
For example, these characteristic times for $\theta=\theta_c=0.50$~rad are
$T_{\rm dec}(\theta_c)\sim10^5$~s and $T_{\rm beam}(\theta_c)\sim10^6$~s.
Moreover, due to the relativistic beaming effect,
such emission is dim.

Using Eq.~(\ref{eq:jet_break}), we calculate the jet break time as
$T_{\rm jet}\sim10^9$~s. 
This large value comes from the large value of $E_0$ and
wideness of the jet ($\theta_j\approx0.6$~rad).
Hence, relatively shallow decay lasts for a long time (see Fig.~\ref{lc-all}).

\subsection{The case of Gaussian jet 2: Gamma-Even-Mass (GEM)}
\label{subsection:GEM}

\begin{figure*}[ht]
\centering 
\includegraphics[width=\linewidth]{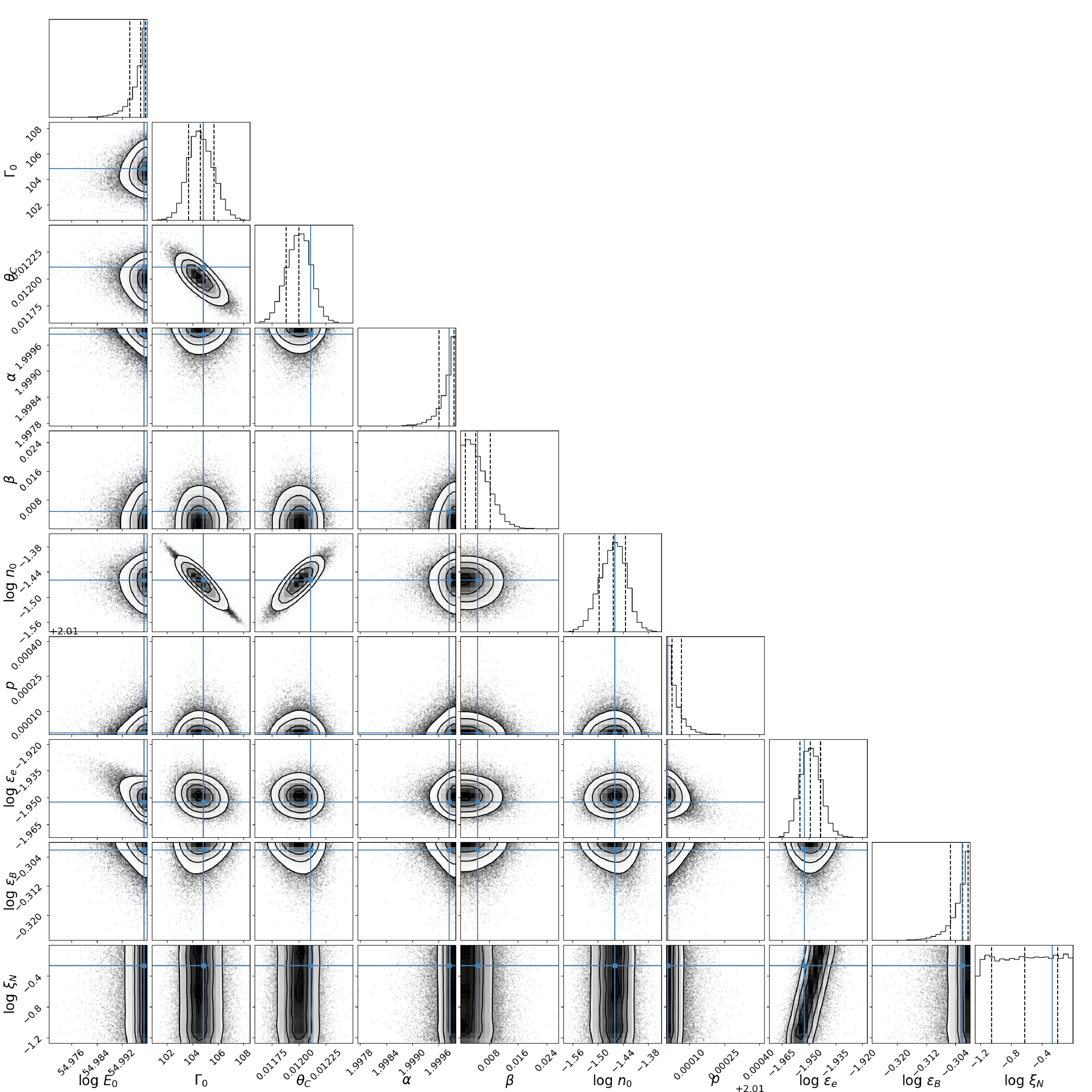}
\caption{
The same as Fig.~\ref{figure:psterior-TH} but for the GEM model.
}
\label{figure:psterior-GEM}
\end{figure*}

Solid lines in Fig.~\ref{lc-all} represent light curves for the best-fit GEM model (Table~\ref{table:best_paramGEM}) 
obtained by MCMC sampling with a prior range $0\leq\beta\leq1.5$.
Figure~\ref{figure:psterior-GEM} shows
the posterior distributions of the model parameters.
In this case, we obtained the smallest value of $\chi^2/{\it d.o.f.}(=10.5)$ in this study,
although the fit is still unacceptable.
We again try to perform fitting with the prior range $1\leq\beta\leq1.5$.
However, in this case, we find that
the initial bulk Lorentz factor at the jet edge satisfies
 $\Gamma(\theta_j)(\theta_j - \theta_{\rm obs})<1$
(see Table~\ref{table:best_param_sum}),
so that the characters of the off-axis afterglow do not appear.
Hence, in this section, we again consider the best-fit model
obtained for the prior range $0\leq\beta\leq1.5$.

As shown in Figs.~\ref{lc-all} and \ref{lc-aroundAP},
this model also gives us optical/infrared achromatic peak at $2\times10^3$~s.
Compared with TH and GF models, GEM model gives
shallower  rising part before the peak,
while steeper decay slope after the peak. This fact
explains the observed late-time r-band and z-band fluxes
at $T\sim3\times10^5$~s.
Hence, reduced chi-squared in this case is the smallest among the model we tried.
Below, we qualitatively explain the behavior of the best-fit model.
%
Observer's line of sight almost coincides with the central axis of the jet
($\theta_{\rm obs}\approx0$).
One obtains $\Gamma(\theta_j)\simeq 15~(<\Gamma_0=1.0\times10^2)$
from inferred the best-fit parameters shown in Table~\ref{table:best_paramGEM}.
The jet is narrower ($\theta_j=0.024$~rad) than the TH and GF models.
Hence, we get $\Gamma(\theta_j)|\theta_{\rm obs} + \theta_j|<1$, so that
the whole jet emission comes to the observer from the beginning.
In such a case, an important timescale for the jet kinematics is $T_{\rm dec}(\theta)$.

Using Eq.~(\ref{eq:Tdec}), we find
$T_{\rm dec}(\theta_{\rm obs}) \sim 10^{3} ~{\rm s}$,
which is on the same order  as the observed peak time $T_{\rm peak}$.
Furthermore, 
it is found from Eqs.~(\ref{eq:nu_m}) and (\ref{eq:nu_c}) that
for $T > 5 \times 10^2~{\rm s}$,
characteristic frequencies satisfy
$\nu_m(T, \theta_{\rm obs}) < \nu_c(T, \theta_{\rm obs}) < \nu_z~(< \nu_r < \nu_X)$.
Then, the observed fluxes in z-band, r-band, and X-ray band 
increases according to $F_\nu \propto T^2$,
and takes maximum simultaneously at $T \sim T_{\rm dec}(\theta_{\rm obs})$.
Compared with the TH and GF cases, the GEM model gives more slowly rising part.
Again, the large value of $E_0 = 1 \times 10^{55}$~erg  and
the small values $n_0 = 2 \times 10^{-2} ~{\rm cm^{-2}}$ and
 $\epsilon_e = 1\times10^{-2}$ are necessary for
$T_{\rm dec}(\theta_{\rm obs}) \sim T_{\rm peak} \sim 10^3$~s
and the observed peak flux of $\sim1$~mJy.
A large value of $\epsilon_B\simeq0.5$ is responsible for efficient radiative energy loss of electrons
to satisfy $\nu_c(T, \theta_{\rm obs}) <\nu_z$ for $T<T_{\rm dec}(\theta_{\rm obs})$.

After $T\sim T_{\rm dec}(\theta_{\rm obs})=2\times10^3$~s,
the emission from angularly separated regions, having the polar angle $\theta$,
arrives at the observer, and it becomes brightest at $T=T_{\rm dec}(\theta)$.
We obtain, from Eqs.~(\ref{gauusianjet}) and (\ref{eq:Tdec}),
$T_{\rm dec}(\theta) \propto \exp{(-\theta^2/6\theta_c^2)}[1 +(\Gamma_0-1)\exp{(-\theta^2/2\theta_c^2)}]^{-8/3}$,
that increases with $\theta$ for $0 \leq \theta \leq \theta_j$.
For our parameters, we get
$T_{\rm dec}(\theta_{\rm obs}\approx0)\sim 10^{3}$~s,
$T_{\rm dec}(\theta_c)\sim 10^{4}$~s, 
and
$T_{\rm dec}(\theta_j)\sim 10^{5}~{\rm s}$.
The maximum flux at $\sim T_{\rm dec}(\theta)$ made by the region with $\theta$  
more rapidly decreases with $\theta$ than in the case of TH jet,
because $E(\theta)$ decreases with $\theta$.
Therefore, for $T>T_{\rm dec}(\theta_c)$ the total observed flux decays steeply.
In this epoch, the decay slope is consistent with
the scaling for jetted afterglow without rapid deceleration due to the sideway expansion in the slow cooling phase
($\nu_m(T, \theta) < \nu_c(T, \theta) < \nu_z < \nu_r < \nu_X$) as
$F_{\nu} \propto T^{-(3p+1)/4}\propto T^{-1.76}$ in z-band, r-band, and X-ray band \citep{panaitescu1998}.
For our parameters, 
$\nu_m(T, \theta_{\rm obs}) < \nu_c(T, \theta_{\rm obs}) < \nu_X$
in the decay phase [see Eqs.~(\ref{eq:nu_m}) and (\ref{eq:nu_c})], so that
the X-ray photon index is predicted as $p/2+1=2.01$ for $p=2.01$,
which is consistent with the observed value of $1.92_{-0.11}^{+0.12}$.

As shown in Fig.~\ref{figure:psterior-GEM},
the parameters $E_0$, $\alpha$ and $\epsilon_B$ reach the margin.
It is unlikely to expand the upper bound of the prior range of $\epsilon_B$ beyond 0.5 $(\log(\epsilon_B) = 10^{-0.3})$.
We attempted the MCMC fitting with extended prior ranges, $50 < \log(E_0) < 57$ and $1 < \alpha < 3$.
In this case,  the best-fit parameters are: 
$E_0 = 1.17 \times 10^{56}~{\rm erg}$, 
$\Gamma_0= 160.70$, 
$\theta_c= 7.64 \times 10^{-3}~{\rm rad}$, 
$\alpha = 2.50$, 
$\beta = 5.95 \times 10^{-3}$, 
$n_0 = 1.29 \times 10^{-2}~{\rm cm^{-3}}$, 
$p = 2.01$, 
$\epsilon_e = 1.01 \times 10^{-3}$, 
$\epsilon_B = 4.82 \times 10^{-1}$, 
and $\xi_N= 0.676$. 
The reduced chi-squared is $923/(96-10)=10.7$, so the fit is not significantly improved. 
The 1D posterior distributions of $E_0$ and $\alpha$ take maximum at
 larger values, $E_0\sim1\times10^{56}$~erg and $\alpha \sim 2.5$, respectively.
Then, the collimation corrected energy 
is calculated as $E_{\rm K,jet} \sim 7\times10^{51}$~erg,
which is too large for the peak energy of the prompt emission $(1+z)E_{\rm peak}\simeq2\times10^2$~keV \citep{zhao2020}. 
Our choice of the prior range, $50 < \log(E_0) < 55$, provides the result that is 
consistent with previous observations for $E_{\rm K,jet}$.

\subsection{Model Comparison}

In any models (TH, GF, and GEM), we need large values of $E_0$ ($\sim10^{54-55}$~erg) to
explain the late-time ($\sim2\times10^3$~s) achromatic peak in the optical/infrared bands.
Even if we impose off-axis viewing $\beta=\theta_{\rm obs}/\theta_j\geq1$ in the fitting,
the results are not drastically improved for the three cases (see Table~\ref{table:best_param_sum}).
The jets with large kinetic energy start to decelerate later than usual, making the achromatic peak.
This seems to best explain the observed result of GRB~080710.
The collimation-corrected energy, $E_{\rm K, jet}$, of the jet in the cases of prior range $0\leq\beta\leq1.5$ are 
calculated
as
$1.2\times 10^{52}$, $1.2\times 10^{54}$, and
$1.3\times 10^{51} ~{\rm erg}$ for the best-fit parameters of TH, GF, and GEM models, respectively.
The GF model provides much larger value than the other events \citep{zhao2020},
which is energetically disfavored.

All six cases 
presented
in this paper gave $\chi^2$/{\it d.o.f.}~$>10$ (see Table~\ref{table:best_param_sum}), 
so that the fit is not acceptable.
However, we see the differences between the $\chi^2$ for TH and GEM models (with prior range $\beta\geq0$)
as $\Delta\chi^2 = \chi^2_{\rm TH} - \chi^2_{\rm GEM} = 1042$, 
and for GF and GEM models (with prior range $\beta\geq0$) as
$\Delta\chi^2 = \chi^2_{\rm GF} - \chi^2_{\rm GEM} = 1224$,
while 
$\Delta$({\it d.o.f})~$=$~({\it d.o.f})$_{\rm TH} -$~({\it d.o.f})$_{\rm GEM} = 1$
and
$\Delta$({\it d.o.f})~$=$~({\it d.o.f})$_{\rm GF} -$~({\it d.o.f})$_{\rm GEM} = 0$.
Using the $\chi^2$-distribution with freedom of unity, we obtain both p-value less than $10^{-15}$.
Hence, the GEM model is the best representing the data among the three models TH, GF, and GEM.

Figure~\ref{AP} shows the r-band to z-band flux ratio, $F_r/F_z$, for six models listed in
Table~\ref{table:best_param_sum}.
One can see that the observed data (black points) is achromatic around the peak time $T_{\rm peak}$.
Thick and thin lines are for cases of different 
prior ranges $0\leq\beta \leq 1.5$ and $1\leq\beta \leq 1.5$, respectively.
Both GEM (red solid lines) and GF (green dashed lines) models are roughly consistent with observation,
explaining the achromatic peak.
The TH (blue dotted lines) models do not fit the data well.
In particular, the TH model with the prior range $1\leq\beta \leq 1.5$ shows the change of the
flux ratio around $T_{\rm peak}$.
In this case ($\theta_{\rm obs}>\theta_j$), 
since initially $\Gamma_0(\theta_{\rm obs}-\theta_j)<1$, the off-axis afterglow character does not appear.
Then, we find the break frequency $\nu_c(T, \theta_j)$ crosses r and z bands in the free expansion phase and 
their crossing times satisfy
$T_c(\theta_j, \nu_r) \lesssim T_c(\theta_j, \nu_z) \sim 2\times 10^{3}$~s.
Hence, the peak times of r and z-band light curves are different,
resulting in the chromatic behavior around $T_{\rm peak}$ for the TH model.

\section{Discussion}
\label{section_discussion}

Using Bayesian inference with MCMC, we have found a possible scenario of observed afterglows of GRB~080710
showing achromatic optical/infrared peak at $T_{\rm peak}=2.2\times10^3$~s after the burst trigger.
One of Gaussian jet models, GEM, best describes the data. 
Since the
 jet has large initial isotropic-equivalent energy ($E_0\sim10^{55}$~erg),
the jet deceleration starts at $T\sim2\times10^3$~s, 
which is responsible for
the observed achromatic peak.
Although $E_0$ is very large, the jet is narrow ($\theta_j=2.4\times10^{-2}$~rad), so that
the value of collimation-corrected jet energy is normal, $E_{\rm K,jet}=1.3\times10^{51}$~erg.
Note that off-axis afterglow interpretation may be unlikely, which is consistent with the observed
bright and hard prompt emission.

Since two observation frequencies $\nu_z$ and $\nu_r$
 are  close to each other, one may expect that the peak is caused by the crossing of $\nu_m(T,\theta)$ to the observed band. 
However, we disfavor this possibility.
For example, the frequency $\nu_m(T,\theta)$ decreases in the adiabatically expanding TH jets through the uniform ambient medium. 
Then, for $\Gamma(r,\theta)|\theta_{\rm obs}-\theta_j|<1$,
we obtain the ratio of the crossing time as $T_m(\theta,\nu_z)/T_m(\theta,\nu_r)=(\nu_z/\nu_r)^{-2/3}\sim1.3$
[see Eq.~(\ref{eq:time_nu_m_on-adi})],
which should be significantly larger than observed.
\citet{kruhler2009} analyzed the unthinned observation data and concluded that the peak is achromatic 
with high measurement accuracy (see their section~3.2.),
that is, $T_m(\theta,\nu_z)/T_m(\theta,\nu_r)\approx1$.
Furthermore, if the peak is made due to the $\nu_m$ crossing, then the flux ratio $F_r/F_z$   becomes larger than unity --- typically, $F_\nu \propto \nu^{1/3}$ for $\nu<\nu_m$, so that one can find $F_r/F_z = (\nu_r/\nu_z)^{1/3} >1$ before the peak time. 
However, as shown in Fig.~\ref{AP}, the observed flux ratio $F_r/F_z$ is always less than unity with a fixed value before and after the peak time. 

Our best-fit GEM model has r-band and z-band light curves with more slowly rising part before the peak time 
compared with
the other models, TH and GF (see Fig.~\ref{lc-aroundAP} and \S~\ref{subsection:GEM}).
Moreover, our GEM model shows more rapid decay after the peak time.
In general, jets in the GEM model have an edge at which initial bulk Lorentz factor
$\Gamma(\theta_j)$ is smaller than that in the central part, $\Gamma(0)$.
Hence, the beaming cone of the emission at the jet edge is wider, so that
the jet edge effects more easily appear for the GEM model.
In particular, for our the best-fit parameters, 
since
the jet is narrowly collimated ($\theta_j=0.02$~rad),
the jet edge can be seen from the observer from the beginning,
resulting in the variation of light-curve shape around and after  the peak time.
In this way, the jet structure in the polar angle direction affects the observed afterglow behavior.
Although the achromaticity around the peak is reproduced, the values of reduced chi-squared 
indicates that the fit is not acceptable.
As shown in Fig.~\ref{lc-aroundAP}, 
the GEM model underpredicts the observed first ($T\approx 4.2\times10^2$~s) 
and the second ($T\approx 5.7\times10^2$~s) data points, although the difference is within a factor of two.
More complicated jet structure might be necessary to better explain the rising part.
Indeed, the hydrodynamics simulations have shown the angular dependence of the
bulk Lorentz factor that is more complicated  than the Gaussian shape
\citep[e.g.,][]{zhangW2003,morsony2007,mizuta2013,gottlieb2021}.
Another possibility for the better fit to the observation 
is to consider the non-uniform circumburst medium like wind profile.
These issues are beyond the scope of this paper, and will be discussed in the future work.

We estimate the efficiency of the prompt emission, 
$\eta_\gamma = E_{\rm \gamma, iso}/[E_{\rm \gamma, iso}+E(\theta_{\rm obs})]$,
where $E(\theta_{\rm obs})$ is the initial kinetic energy of the afterglow jet
along the observer's line of sight.
For the three cases, TH, GF, and GEM, we found
$E(\theta_{\rm obs}) \sim 10^{54-55} ~{\rm erg}$, so that we obtain
$\eta_\gamma \sim 10^{-4}$--$10^{-3}$.
This value is much smaller than inferred values so far
\citep[e.g.,][]{lloyd2004,beniamini2016}.
The jet of GRB~080710 has large energy but the radiation efficiency of the prompt emission is small.
A possible reason for the low prompt efficiency is less turbulent flow in the GRB jet
--- for example, small velocity difference in the radial direction --- 
resulting in the inefficient internal dissipation. 

Since the deceleration time scales as $T_{\rm dec}\propto \Gamma_0^{-8/3}$,
one may expect that the late-time achromatic peak arises for baryon-loaded jets (dirty fireballs) with initial 
bulk Lorentz factor of around 30.
Taking into account this possibility, we have taken the
flat prior with a range $10\leq\Gamma_0\leq500$, and searched for the best-fit parameters.
Then, we have obtained the posterior distribution indicating $\Gamma_0\gtrsim100$.
For the best-fit parameters, one can find $\Gamma(\theta_{\rm obs})\gtrsim100$.
In practice,  if we take $\Gamma_0=30$--50, $E_0 \sim 10^{52} ~{\rm erg}$, and $\theta_{\rm obs}=0$ 
in order for the jet to have $T_{\rm dec}\sim 10^{3} ~{\rm s}$, then 
the peak flux becomes much smaller than observed.
The brightness at the peak time ($\sim1$~mJy at infrared/optical) is also a strict constraint on the modeling.

As introduced in Section~1,
GRB~080710 had the prompt emission properties, 
$E_{\gamma, \rm{iso}}=6\times10^{51}$~erg, $(1+z)E_{\rm peak}\simeq2\times10^2$~keV and
$L_{\gamma,{\rm iso}}=1-8\times10^{50}$~erg~s$^{-1}$, 
and we have found, in this paper, the initial bulk Lorentz factor of the afterglow jet,
$\Gamma_0\approx1-2\times10^2$.
It is believed to be the same as bulk Lorentz factor of the emitting region of the prompt emission.
These are within typical range of values
\citep[e.g.,][]{liang2010,liang2015,lu2012,ghirlanda2018}.

Our result indicates that GRB~080710 occurred in the rarefied medium ($n_0\sim10^{-2}$~cm$^{-3}$).
Recent bright event GRB~221009A might also arise in the small-density region
\citep[e.g.,][]{sato2023}.
Such events may appear in bubbles made by strong wind from progenitor stars
or in superbubbles made by OB association.
Furthermore, existence of the low-density region has also been identified
by three-dimensional magnetohydrodynamics simulations for a part of interstellar medium
of star forming galaxies \citep{deAvillez2005}.

We also attempted the MCMC fitting with the power-law jet model, $E(\theta)= E_0(1+\theta^2/b\theta_c^2)^{-b/2}$ for $\theta<\theta_j$, using the same conditions as those of the Gaussian-GEM model.
Then, we obtained the best-fit model with reduced chi-squared of $782/(96-11)=9.2$.
The fit is not significantly improved, although this value is slightly lower than that for the Gaussian-GEM model of $10.5$. 
This result is  reasonable because the functional forms of both power-law and Gaussian models are identical, $E(\theta)\sim E_0 (1-\theta^2/2\theta_c^2)$, for small $\theta$. 
Hence, our basic claim for GRB~080710 afterglow 
--- that is,
a narrow, structured jet with $E_0\sim10^{55}$~erg ---
remains unchanged even if the power-law jets are considered.

Our results indicate that there is a diversity of GRB jets, in particular, 
the existence of narrowly collimated outflows with large isotropic equivalent kinetic energy
but very low efficiency of the prompt emission.
However, such GRB events may be difficult to be triggered since their solid angle 
 is smaller than that of typical events so far. 
The prompt emission from  misaligned jets is dim but
their external shock emission might be detectable as orphan afterglows.
Event rate of such phenomena is difficult to be calculated since we do not precisely know 
the fraction of the narrow jets with the large isotropic kinetic energy.
The population of narrow structured jets may be studied through the other
events (GRBs~050408, 071031, 080603A) that possess similar afterglow properties to GRB~080710, 
which will be done in future work.
The multi-color follow-up observations  are important 
to study if observed afterglow peak is achromatic or not.

In this paper, we propose the large isotropic equivalent kinetic energy of the jet 
($E_0\sim10^{55}$~erg) for
GRB~080710 afterglow, while the small opening angle of the jet $\theta_j$ makes
the collimation-corrected jet energy $E_{\rm K,jet}$  normal for TH and GEM cases.
This situation may be similar to the case of GRB~221009A
\citep{laskar2023,sato2023,lhaaso2023}.
However, in the case of GRB~080710, the energy of the prompt gamma-ray emission
is much smaller (i.e., $\eta_\gamma\ll 1)$, while for GRB~221009A  
the isotropic equivalent gamma-ray and afterglow kinetic energies are comparable.
Such kind of ``monster'' events like GRB~080710 with low gamma-ray emission efficiency might 
release their energy in the different form, that is, they might
be the source of ultra-high-energy cosmic rays and/or neutrinos.

\section*{Acknowledgements}
We thank 
Kento~Aihara,
Kazuyoshi~Tanaka,
Drs.
Tomoki~Matsuoka,
Tsuyoshi~Inoue,
Motoko~Serino,
Makoto~Uemura
and
Koji~S.~Kawabata
for valuable comments,
and also thank
Drs.~G.~Ryan
and H.~van~Eerten
for kind instructions on \texttt{afterglowpy}.
We are grateful to the anonymous referee for his or her comments to improve the paper.
This work was supported by JST SPRING, Grant Number JPMJSP2103 (KO).
This research
was partially supported by JSPS KAKENHI Grant Nos.~22KJ2643 (YS), 
JP20KK0064 (SJT), JPJSBP120229940 (SJT),
22H01251 (RY) 23H01211 (RY), 23H04899 (RY),
23H04895 (TS), and 23H01216 (TS),
by the Sumitomo Foundation (SJT) and by the Research Foundation For Opto-Science and Technology (SJT).


\appendix
\section{Analytical estimate of characteristic times and frequencies, and the peak flux}
\label{appendix_dynamics}

In this Appendix, we present analytical formulae for
characteristic timescales for an observer,
$T_{\rm dec}(\theta)$,
$T_{\rm beam}(\theta)$,
$T_{\rm jet}$,
$T_m(\theta,\nu)$, and $T_c(\theta,\nu)$,
where
$\theta$ is a polar angle from the central axis of the jet ($\theta=0$), and
$\nu$ is the observation frequency.
Characteristic frequencies, 
$\nu_m(T,\theta)$ and $\nu_c(T,\theta)$,
at the observer time $T$
for the synchrotron emission from the region at $(r,\theta)$ are also given.
In order to analyze the results for structured jet models,
we provide expressions by considering 
the angular dependence.
As defined in Section~\ref{section-title:modeling},
the initial angular distributions of the isotropic equivalent energy and 
the bulk Lorentz factor are denoted by $E(\theta)$ and $\Gamma(\theta)$, respectively.
Note that $E(0) = E_0$ and $\Gamma(0) = \Gamma_0$.

Let $\Gamma(r, \theta)$ be the bulk Lorentz factor of emitting material
that is radially moving in the direction $\theta$ and at the radial coordinate $r$ from the central engine located at $r=0$.
The Lorentz factor is  measured in the rest frame, $(r,\theta)$, of the central engine.
In the initial coasting phase, the bulk Lorentz factor is constant with $r$, so that
\begin{align}\label{eq:Gamma_free}
\Gamma(r, \theta) = \Gamma(0, \theta) = \Gamma(\theta).
\end{align}
When the material sweeps enough surrounding matter, the free expansion phase ends, and the jet enters into the adiabatic deceleration phase 
during which the dynamical evolution of the emitting material is given by
\begin{align}\label{eq:r_theta}
  r(\theta)
  \sim \left( \frac{3E(\theta)}{4 \pi n_0 m_p c^2 \Gamma(r, \theta)^2} \right)^{\frac{1}{3}},
\end{align}
where we assume a uniform ambient matter with the number density $n_0$, 
and $m_p$ and $c$ are proton mass and the speed of light, respectively.
Substituting  $\Gamma(r,\theta)=\Gamma(\theta)$ into Eq.~(\ref{eq:r_theta}),
we define the deceleration radius, $R_{\rm dec}(\theta)$, around which the material 
starts to decelerate, 
\begin{align}\label{eq:Rdec}
  R_{\rm dec}(\theta) 
  = \left( \frac{3E(\theta)}{4 \pi n_0 m_p c^2 \Gamma(\theta)^2} \right)^{\frac{1}{3}}.
\end{align}
If the emitting material is initially far from the observer in the direction $\theta_{\rm obs}$
in a sense, $\Gamma(\theta)|\theta_{\rm obs}-\theta|>1$, 
then the emission is initially dim for the observer.
As the emitter decelerates, the angle of the relativistic beaming cone $\Gamma(r,\theta)^{-1}$ increases.
The beaming radius, $R_{\rm beam}(\theta)$, at which
$\Gamma(r,\theta)|\theta_{\rm obs}-\theta|=1$ is given by
\begin{align}\label{eq:Rbeam}
  R_{\rm beam}(\theta) 
  &= \left( \frac{3 E(\theta)}{4 \pi n_0 m_p c^2} \right)^{\frac{1}{3}} |\theta_{\rm obs}-\theta|^{\frac{2}{3}}.
\end{align}
The observed flux from the emitting material with $\theta$ at $r>R_{\rm beam}(\theta)$ is almost
indistinguishable from that by an emitter going toward the observer ($\theta=\theta_{\rm obs})$.
Now, we define the observer time $T(\theta)$ at which photons from the emitter  moving into $\theta$ 
arrive at the observer, that is given by
\begin{align}\label{eq:t_rtheta}
  T(\theta)
  &\equiv T(r(\theta), \theta)\notag\\
  &= (1+z)\frac{r(\theta)[1 - \beta(r,\theta) \cos(\theta_{\rm obs}-\theta)]}{\beta(r,\theta) c} \notag\\
  &\approx (1+z)\frac{r(\theta)}{2c\Gamma(r,\theta)^2}(1 + \Gamma(r, \theta)^2 |\theta_{\rm obs}-\theta|^2),
\end{align}
where $z$ and $\beta(r, \theta) = v(r, \theta)/c$ are the redshift of the source and
the velocity of the emitter measured in the central engine frame, respectively,
and
$T=0$ is defined as the time at which a photon simultaneously emitted by the central engine
with the jet arrives at the observer.
Hence, $T_{\rm dec}(\theta)$ and $T_{\rm beam}(\theta)$ are calculated as
\begin{align}\label{eq:Tdec}
    T_{\rm dec}(\theta)
    &= T(R_{\rm dec}(\theta), \theta)\notag\\
    &\sim (1+z)\left( \frac{3E(\theta)}{32 \pi n_0 m_p c^5} \right)^{\frac{1}{3}}\times
    \left\{ \begin{array}{ll}
        \Gamma(\theta)^{-\frac{8}{3}}& ({\rm for}~\Gamma(\theta)|\theta_{\rm obs}-\theta| < 1) \\
        |\theta_{\rm obs}-\theta|^2\Gamma(\theta)^{-\frac{2}{3}} & ({\rm for}~\Gamma(\theta)|\theta_{\rm obs}-\theta| > 1)
    \end{array} \right.~~,
\end{align}
and
\begin{align}\label{eq:beam_time}
  T_{\rm beam}(\theta) 
  &= T(R_{\rm beam}(\theta), \theta)\notag\\
  &\sim (1+z)\left( \frac{3 E(\theta)}{32 \pi n_0 m_p c^5} \right)^{\frac{1}{3}} \left|\theta_{\rm obs}-\theta \right|^{\frac{8}{3}},
\end{align}
respectively.
If the jet is initially fat in a sense, $\Gamma(\theta_j)(\theta_{\rm obs}+\theta_j)>1$, then, the jet edge cannot be seen at the beginning. As the jet decelerates, the beaming angle becomes large.
When $\Gamma(r, \theta_j)=|\theta_{\rm obs} + \theta_j|^{-1}$, 
the whole jet emission becomes visible. This jet break time is given by,
\begin{align}\label{eq:jet_break}
  T_{\rm jet}
  \sim (1+z)\left( \frac{3 E(\theta_j)}{32 \pi n_0 m_p c^5} \right)^{\frac{1}{3}} \left| \theta_{\rm obs} + \theta_j \right|^{\frac{8}{3}}.
\end{align}

Let $\nu'_{\rm syn}(\gamma)$ be the characteristic frequency of 
the synchrotron emission from an electron with Lorentz factor $\gamma$ measured in the comoving frame of the emitter.
The observed spectrum of the synchrotron emission from a radially moving material
in the direction $\theta$ that contains electrons following the power-law
energy distribution with an index $p$ and the minimum Lorentz factor $\gamma_m(\theta)$
has a break at the frequency given by
\begin{align}\label{eq:nu_m}
  \nu_m(T,\theta)
  & = \frac{\nu'_{\rm{syn}}\left( \gamma_m(\theta) \right) \delta(r,\theta)}{1+z}\notag\\
  &= \frac{2 \sqrt{2} e m_p^{5/2}}{(1+z)\pi^{1/2} m_e^3} n_0^{1/2} \epsilon_B^{1/2} \epsilon_e^2 \xi_N^{-2}\left( \frac{p-2}{p-1} \right)^2 \frac{2\Gamma(r, \theta)^4}{1+\Gamma(r, \theta)^2 |\theta_{\rm obs} - \theta|^2},
\end{align}
and the cooling frequency given by
\begin{align}\label{eq:nu_c}
  \nu_c(T,\theta)
  &= \frac{\nu'_{\rm{syn}}(\gamma_c(\theta)) \delta(r, \theta)}{1+z}\notag\\
  &= \frac{9 e \sqrt{2} m_e (1+z)}{128 \pi^{1/2} \sigma_T^2 c^2 m_p^{3/2} n_0^{3/2} \epsilon_B^{3/2}} 
  \frac{2[1+\Gamma(r,\theta)^2 |\theta_{\rm obs} - \theta|^2]}{\Gamma(r,\theta)^4}\frac{1}{T(\theta)^2},
\end{align}
where electrons with Lorentz factor greater than $\gamma_c(\theta)$ cools efficiently, and
$\delta(r,\theta)=1/\Gamma(r,\theta)[1-\beta(r,\theta)\cos(\theta_{\rm obs}-\theta)]$ is the Doppler factor.
Note that Eqs.~(\ref{eq:nu_m}) and (\ref{eq:nu_c}) contain radial coordinate $r$ , which can be rewritten
by $T$ and $\theta$, that is, $r=r(T,\theta)$ [see Eqs.~(\ref{eq:r_theta}) and/or (\ref{eq:t_rtheta})].
Hence, both $\nu_m$ and $\nu_c$ can be treated as functions of $T$ and $\theta$.
The observer sees that $\nu_m(T, \theta)$ and $\nu_c(T, \theta)$ cross the observation frequency $\nu$
at $T_m(\theta, \nu)$ and $T_c(\theta, \nu)$, respectively.
In the free expansion phase, since $\nu_m(T, \theta)$ is time-independent, $T_m(\theta, \nu)$ is undefined, while
$T_c(\theta, \nu)$ is approximated for $\Gamma(r,\theta)|\theta_{\rm obs} - \theta_j|<1$ as
\begin{align}\label{eq:time_nu_c_on-free}
  T_c(\theta, \nu)
  &\simeq (1+z)^{1/2}\frac{3 \sqrt{2} e^{1/2} m_e^{1/2}}{16 \pi^{1/4} m_p^{3/4}\sigma_T c \Gamma(\theta)^2} n_0^{-3/4} \epsilon_B^{-3/4} \nu^{-1/2}.
\end{align}
In the adiabatic deceleration phase, approximated formulae of
$T_m(\theta, \nu)$ and $T_c(\theta, \nu)$ for $\Gamma(r,\theta)|\theta_{\rm obs} - \theta_j|<1$ are given by 
\begin{align}\label{eq:time_nu_m_on-adi}
  T_m(\theta, \nu)
  &\simeq (1+z)^{1/3}\frac{6^{1/3} e^{2/3} m_p^{4/3}}{4 \pi^{2/3} m_e^2 c^{5/3}} \left( \frac{p-2}{p-1} \right)^{4/3} \epsilon_B^{1/3} \epsilon_e^{4/3} \xi_N^{-4/3} E(\theta)^{1/3} \nu^{-2/3},
\end{align}
and
\begin{align}\label{eq:time_nu_c_on-adi}
  T_c(\theta, \nu)
  &\simeq (1+z)^{-1}\frac{27 e^2 m_e^2 c}{32 m_p^2 \sigma_T^4} n_0^{-2} \epsilon_B^{-3} E(\theta)^{-1} \nu^{-2},
\end{align}
respectively.

\newpage
\begingroup
\renewcommand{\arraystretch}{1.3}
  \begin{table}[H]
    \caption{
    Parameter estimation of the top-hat (TH) jet describing GRB~080710 afterglow.
    Prior ranges for each parameters are shown in the second column.
    We adopt a uniform distribution for each prior.
    Here, we define $\beta = \theta_{\rm obs}/\theta_j$.
    The median values of one-dimensional (1D) posterior distributions of each parameter are presented (third column) with the symmetric 68\% uncertainties (i.e. the 16\% and 84\% quantiles).
    We show, in the right-most column, the best-fit model parameters that gives us 
    the maximum posterior probability among all steps.
    For the two parameter sets, the values of $\chi^2$ and reduced chi-squared for 96 data points are shown.}
    \label{table:best_paramTH}
    \centering
    \begin{tabular}{cccc}
      \hline
      Parameters & prior ranges & 1D dist. & best-fit \\
      \hline
      $\theta_j$ & $[0.001, 0.5]$ & $0.0821^{+0.0062}_{-0.0076}$ &  $0.0719$\\
      $\beta$& $[0, 1.5]$ & $0.762^{+0.0002}_{-0.0001}$ & $0.762$\\
      $\log_{10}(E_0 ~[{\rm erg}])$ & $[50, 55]$ & $54.60^{+0.28}_{-0.41}$ &  $54.61$\\
      $\log_{10}(n_0 ~[{\rm cm^{-3}}])$ & $[-4, 2]$ & $-1.104^{+0.422}_{-0.470}$ & $-1.489$\\
      $\Gamma_0$ & $[10, 500]$ & $97.28^{+9.73}_{-6.71}$ & $110.98$\\
      $p$ & $[2.01, 2.9]$ & $2.478^{+0.004}_{-0.004}$ & $2.48$\\
      $\log_{10} \epsilon_e$ & $[-3, -0.3]$ & $-2.52^{+0.34}_{-0.25}$ & $-2.60$\\
      $\log_{10} \epsilon_B$ & $[-5, -0.3]$ & $-2.33^{+0.44}_{-0.36}$ & $-2.09$\\
      $\log_{10} \xi_N$ & $[-2, 0]$ & $-0.395^{+0.283}_{-0.401}$ & $-0.609$\\
      \hline
      $\chi^2$/{\it d.o.f} &  & $31.5$ & $22.3$\\
      $\chi^2$ &   & $2737$ & $1942$\\
      {\it d.o.f} &  & $96-9$ & $96-9$\\
      \hline
    \end{tabular}
  \end{table}
\endgroup

\begingroup
\renewcommand{\arraystretch}{1.3}
  \begin{table}[H]
    \caption{
    The same as Table~\ref{table:best_paramTH} but for Gaussian Gamma-Flat (GF) model. We further define $\alpha = \theta_j/\theta_c$.
    }
    \label{table:best_paramGF}
    \centering
    \begin{tabular}{cccc}
      \hline
      Parameters & prior ranges & 1D dist. & best-fit \\
      \hline
      $\theta_c$ & $[0.001, 0.5]$ & $0.4994^{+0.0004}_{-0.0009}$ &  $0.49991$\\
      $\alpha$& $[1, 2]$ & $1.191^{+0.007}_{-0.011}$ & $1.18$\\
      $\beta$& $[0, 1.5]$ & $1.007^{+0.0001}_{-0.00007}$ & $1.01$\\
      $\log_{10}(E_0 ~[{\rm erg}])$ & $[50, 55]$ & $54.999^{+0.0005}_{-0.0011}$ &  $54.998$\\
      $\log_{10}(n_0 ~[{\rm cm^{-3}}])$ & $[-4, 2]$ & $-1.679^{+0.034}_{-0.021}$ & $-1.65$\\
      $\Gamma_0$ & $[10, 500]$ & $162.81^{+1.92}_{-1.75}$ & $162.11$\\
      $p$ & $[2.01, 2.9]$ & $2.767^{+0.002}_{-0.002}$ & $2.77$\\
      $\log_{10} \epsilon_e$ & $[-3, -0.3]$ & $-2.015^{+0.004}_{-0.005}$ & $-2.02$\\
      $\log_{10} \epsilon_B$ & $[-5, -0.3]$ & $-2.478^{+0.013}_{-0.023}$ & $-2.49$\\
      $\log_{10} \xi_N$ & $[-2, 0]$ & $-6.5^{+4.8}_{-11.2} \times 10^{-4}$ & $-0.001$\\
      \hline
      $\chi^2$/{\it d.o.f} &  & $24.4$ & $24.7$\\
      $\chi^2$ &   & $2095$ & $2124$\\
      {\it d.o.f} &  & $96-10$ & $96-10$\\
      \hline
    \end{tabular}
  \end{table}
\endgroup

\clearpage
\begingroup
\renewcommand{\arraystretch}{1.3}
  \begin{table}[ht]
    \caption{
    The same as Table~\ref{table:best_paramGF} but for Gaussian Gamma-Even-Mass (GEM) model.
    }
    \label{table:best_paramGEM}
    \centering
    \begin{tabular}{cccc}
      \hline
      Parameters & prior ranges & 1D dist. & best-fit \\
      \hline
      $\theta_c$ & $[0.001, 0.5]$ & $0.0120^{+0.0001}_{-0.0001}$ &  $0.0121$\\
      $\alpha$& $[1, 2]$ & $1.9999^{+0.0001}_{-0.0002}$ & $1.9998$\\
      $\beta$& $[0, 1.5]$ & $0.00418^{+0.00422}_{-0.00281}$ & $0.00474$\\
      $\log_{10}(E_0 ~[{\rm erg}])$ & $[50, 55]$ & $54.998^{+0.002}_{-0.003}$ &  $54.9990$\\
      $\log_{10}(n_0 ~[{\rm cm^{-3}}])$ & $[-4, 2]$ & $-1.463^{+0.029}_{-0.034}$ & $-1.46$\\
      $\Gamma_0$ & $[10, 500]$ & $104.59^{+1.07}_{-0.92}$ & $104.84$\\
      $p$ & $[2.01, 2.9]$ & $2.01^{+0.00004}_{-0.00002}$ & $2.01$\\
      $\log_{10} \epsilon_e$ & $[-3, -0.3]$ & $-1.949^{+0.006}_{-0.006}$ & $-1.95$\\
      $\log_{10} \epsilon_B$ & $[-5, -0.3]$ & $-0.302^{+0.002}_{-0.003}$ & $-0.302$\\
      $\log_{10} \xi_N$ & $[-2, 0]$ & $-0.622^{+0.423}_{-0.432}$ & $-0.265$\\
      \hline
      $\chi^2$/{\it d.o.f} &  & $10.9$ & $10.5$\\
      $\chi^2$ &   & $935$ & $900$\\
      {\it d.o.f} &  & $96-10$ & $96-10$\\
      \hline
    \end{tabular}
  \end{table}
\endgroup

\begingroup
\renewcommand{\arraystretch}{1.3}
  \begin{table}[ht]
    \caption{
    The best-fit parameters describing GRB~080710 afterglow for six models:
    Top-hat (TH), Gaussian Gamma-Flat (GF), Gaussian Gamma-Even-Mass (GEM), 
    TH ($\beta\geq1$), GF ($\beta\geq1$), and GEM ($\beta\geq1$).
    In the last three models, the prior range of $\beta=\theta_{obs}/\theta_j$ is limited to $1\leq\beta\leq1.5$.
    The best-fit parameters are the values when the posterior probability is largest.
    The values of $\chi^2$ and reduced chi-squared for 96 data points are also shown in this table.
    }
    \label{table:best_param_sum}
    \centering
    \begin{tabular}{ccccccc}
      \hline
      \shortstack{Model} & \shortstack{TH\\ ($\beta \geq 0$)} & \shortstack{GF\\ ($\beta \geq 0$)} &  \shortstack{GEM\\ ($\beta \geq 0$)} & \shortstack{TH\\ ($\beta \geq 1$)} & \shortstack{GF\\ ($\beta \geq 1$)} &  \shortstack{GEM\\ ($\beta \geq 1$)}\rule[0pt]{0pt}{5pt}\\
      \hline \hline
      $\theta_{\rm c} [\rm{rad}]$ & $\cdots$  & $5.00 \times 10^{-1}$ & $1.21 \times 10^{-2}$ & $\cdots$ & $5.00 \times 10^{-1}$ & $1.07 \times 10^{-2}$ \\
      $\theta_{\rm j} [\rm{rad}]$ & $7.19 \times 10^{-2}$ & $5.92 \times 10^{-1}$ & $2.42 \times 10^{-2}$ & $6.71 \times 10^{-3}$ & $5.98 \times 10^{-1}$ & $1.07 \times 10^{-2}$ \\
      $\theta_{\rm{obs}} [\rm{rad}]$ & $5.48 \times 10^{-2}$  & $5.96 \times 10^{-1}$ & $1.15 \times 10^{-4}$ & $6.72 \times 10^{-3}$ & $6.02 \times 10^{-1}$ & $1.07 \times 10^{-2}$ \\
      $E_0[\rm{erg}]$  & $4.09 \times 10^{54}$ & $9.95 \times 10^{54}$ & $9.98 \times 10^{54}$ & $9.95 \times 10^{54}$ & $9.99 \times 10^{54}$ & $9.99 \times 10^{54}$ \\
      $n_0 [\rm{cm^{-3}}]$ & $3.24 \times 10^{-2}$ & $2.23 \times 10^{-2}$ & $3.47 \times 10^{-2}$ & $6.89 \times 10^{-4}$ & $2.33 \times 10^{-2}$ & $2.59 \times 10^{-2}$ \\
      $\Gamma_0$ & $1.11 \times 10^{2}$ & $1.62 \times 10^{2}$ & $1.04 \times 10^{2}$ & $2.09 \times 10^{2}$ & $1.62 \times 10^{2}$ & $1.49 \times 10^{2}$ \\
      $p$ & $2.48$ & $2.77$ & $2.01$ & $2.01$ & $2.77$ & $2.01$ \\
      $\epsilon_e$ & $2.50 \times 10^{-3}$ & $9.65 \times 10^{-3}$ & $1.12 \times 10^{-2}$ & $1.49 \times 10^{-2}$ & $9.7 \times 10^{-3}$ & $1.9 \times 10^{-2}$ \\
      $\epsilon_B$ & $8.13 \times 10^{-3}$ & $3.24 \times 10^{-3}$ & $4.99 \times 10^{-1}$ & $5.00 \times 10^{-1}$ & $3.1 \times 10^{-3}$ & $4.98 \times 10^{-1}$ \\
      $\xi_N$ & $0.246$ & $0.997$ & $0.543$ & $0.177$ & $0.996$ & $0.648$ \\
      \hline
      $\chi^2$/{\it d.o.f} & $22.3$  & $24.7$ & $10.5$ & $18.3$ & $24.3$ & $12.1$ \\
      $\chi^2$ & $1942$  & $2124$ & $900$ & $1589$ & $2091$ & $1037$ \\
      {\it d.o.f} & $96-9$  & $96-10$ & $96-10$ & $96-9$ & $96-10$ & $96-10$ \\
      \hline
    \end{tabular}
  \end{table}
\endgroup

\clearpage

\clearpage

\clearpage

\clearpage

\end{document}